# Multiphysics Simulation Framework for Continuous Ion Transport and Energetics in Atmospheric Pressure Interfaces

Yihui Yan[1], Jozef Lengyel[2], Frédéric Rosu[1], and Valérie Gabelica[1*]

[1]*School of Pharmaceutical Sciences, University of Geneva, 1205 Geneva, Switzerland*

[2]*School of Natural Sciences, Technical University of Munich, 85748 Garching, Germany.*

[*] Corresponding Author: Valérie Gabelica (Valerie.Gabelica@unige.ch)

**ABSTRACT.** Optimizing atmospheric pressure interfaces requires balancing efficient ion transport with unwanted collision activation. The complex environment at the inlet region, including rapid pressure drops, RF/DC electric fields, ion-gas collisions, and space charge effects, makes simulation of this system challenging. In this work, we introduce MuCITE (Multiphysics simulation for Continuous Ion Transport and Energetics), which combines continuous ion injection, precomputed electric and gas flow fields, and particle-in-cell space charge updates. Furthermore, by employing improved impulsive collision theory, it can not only predict transport but also analyze the loss position of ions and the evolution of their kinetic energy, internal energy, and dissociation probability during transport. Transport predictions with ion currents and activation trends of in-source fragmentation were compared with experimental measurements using a custom-built source and a Q-Exactive HF source. The results show that the simulation predictions agree well with the experimental results and can serve as a reference before experiments. MuCITE allows users to derive ion transport, output beam characteristics, and ion energy through the configurable interface geometry and operating conditions. It supports simulation-guided design and tunable parameters to achieve predicted efficient ion transport or in source activation.

Keywords: ion trajectory; space charge; internal energy; thermodynamic properties

## INTRODUCTION

Electrospray ionization (ESI) enables the transfer of fragile and high mass ions from solution into the gas phase. However, a substantial fraction of the generated ions is lost before reaching the mass analyzer [1, 2]. In mass spectrometers, ions are typically sampled from ambient conditions through a capillary or aperture and must subsequently be confined and transported into lower pressure stages. This region, commonly referred to as the atmospheric pressure ionization (API) source, usually employs DC and RF ion optics for efficient ion transfer. The design and configuration of the API directly influence the measured ion current, charge state distribution, and internal energy of transmitted ions. This is important when high sensitivity must be balanced against soft ion transfer to preserve fragile ions and noncovalent complexes [3].

A wide range of API source components, including capillary inlets, skimmers, multipole ion guides, and stacked ring ion guides such as ion funnels and S-lens, have been developed to improve ion transfer during the rapid gas expansion [4-6]. Studies addressing API source performance have shown that realistic gas fields, electrode geometry, and space charge effects must be considered together [7-10].

Numerical trajectory modeling offers a practical route to separate coupled transport mechanisms before new source hardware is built [11-13]. SIMION remains a standard tool for ion optical field calculation and trajectory simulations [14, 15]. COMSOL Multiphysics provides a general multiphysics environment supporting charged particles transfer, gas collisions, fluid flow, and bidirectional particle field coupling [16]. Recent frameworks such as SimELIT [17], IDSimF [18-20], ICARION [21] provide reusable infrastructure for ion trajectory, molecular ion dynamics, external fields, space charge, and collision models. IonSPA, based on an improved impulsive collision theory (IICT), provides a complementary description of collision mediated translational energy damping and molecular heating or cooling [22-24]. SimIonSPA integrates this collision treatment into SIMION, enabling trajectory-resolved analysis, including the recording of electrode-impact "splat" events [25-28]. What is still missing is a unified workflow that combines these elements: i.e., a continuous injection of large load of ions, electrical fields, gas flow, current dependent Particle-in-Cell (PIC) space charge, spatial loss classification, and internal energy prediction in one reproducible simulation path.

Multiphysics simulation for Continuous Ion Transport and Energetics (MuCITE) addresses this need as a modular continuous ion transfer framework. SIMION potential arrays, prescribed Computational Fluid Dynamics (CFD) fields, electrode masks, ion properties, and operating conditions are supplied as inputs. The common runtime returns weighted ion transmission, electrode loss maps, exit beam dimensions, kinetic energy distributions, effective internal temperature distributions, and model dependent precursor dissociation propensity. The two S-lens equipped API source studied here serve as representative verification and application cases for tracing ion transmission, identifying geometric loss regions, preserving low activation for soft or native transfer, or increasing activation for in-source collisional induced dissociation (CID).

## METHODS

MuCITE is developed in Python 3.11 and adapts Taichi 1.7.4 kernel to perform particle propagation, field interpolation, PIC charge deposition, and adaptive time-step estimation. Sparse Poisson equations are solved using SciPy and PyAMG; additional open-source library dependencies include NumPy, pandas, Matplotlib, and Pillow. Windows is the formally tested platform. Linux and macOS are supported. MuCITE is released under the GNU General Public License version 3.

### Simulation scope and modular architecture.

Figure 1 summarizes the MuCITE workflow. Offline preprocessing aligns externally calculated DC and RF potentials, prescribed gas flow fields, and an electrode mask on a common axisymmetric $(r, z)$ coordinate system, as called "field baking" in MuCITE. An ion template specifies mass, charge state, collision cross section, atom count, heat capacity profile, and optional activation parameters. A separate source definition specifies current, spatial profile, injection plane, and initial velocity distribution. During runtime, a defined number of real ions are packaged into weighted ion packs and propagated in three dimensions, with field interpolation, space charge updates, collision handling, and terminal event classification implemented as common modules shared across source geometries.

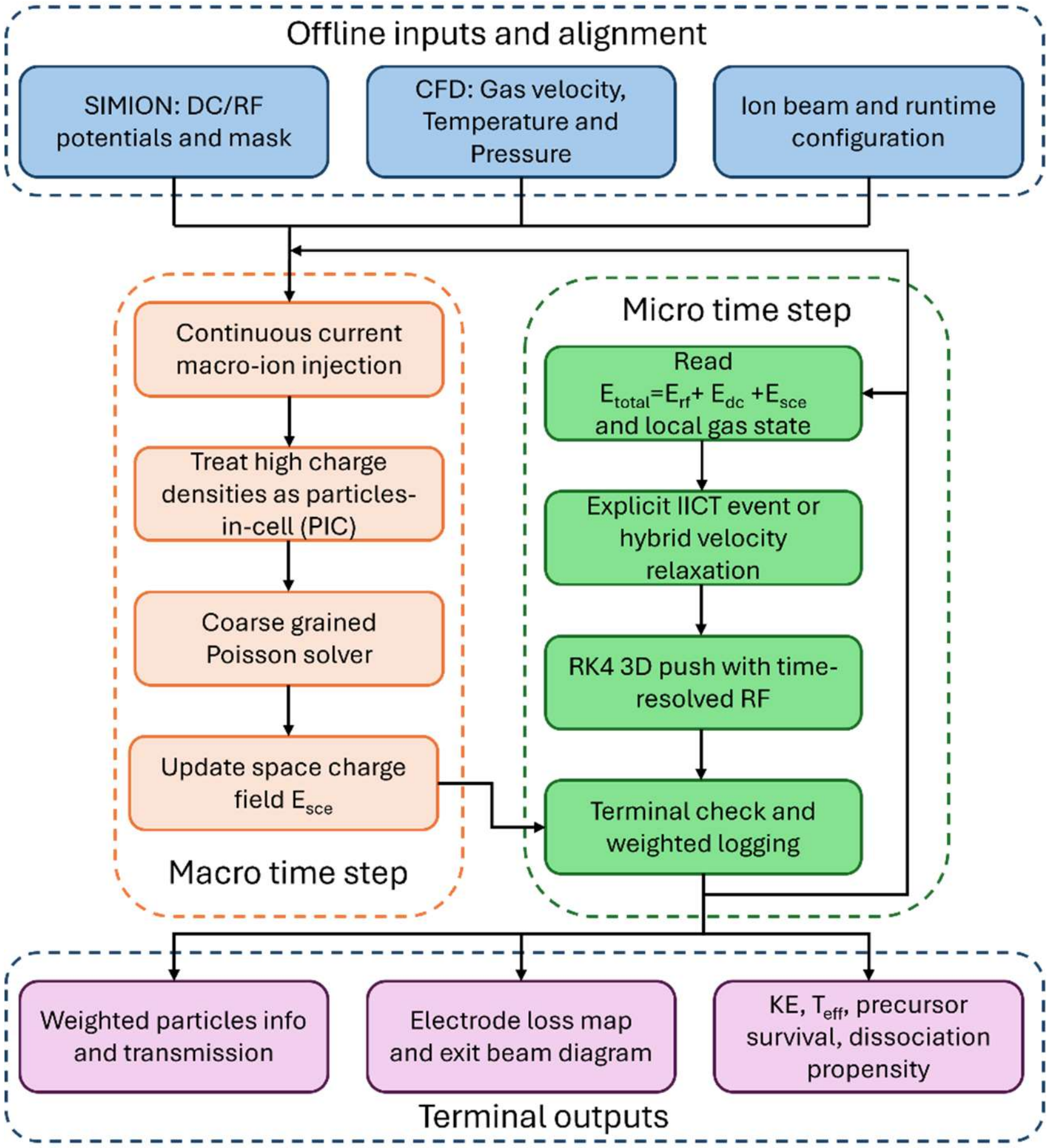

**Figure 1.** MuCITE workflow.

The runtime uses operator splitting to separate PIC field updates from particle transport. At each macro step, charge carried by the active three-dimensional weighted ion packs is deposited onto an axisymmetric PIC grid. The space charge Poisson equation is then solved, and the resulting field is coupled with the precomputed DC and RF fields. While the space charge field remains fixed, micro steps update collisions or Langevin relaxation, evaluate the time dependent RF field, advance particle positions and velocities using $4^{th}$-order Runge–Kutta (RK4) integration, and classify terminal events.

At each particle push, a global micro step size is selected as

$$\Delta t_{global} = min(0.05t_{RF}, 0.1\tau_{collision}, \Delta t_{acc}, \Delta t_{CFL}, \Delta t_L, \Delta t_{rem}) \quad (1)$$

where the $0.05t_{\mathrm{RF}}$ resolves each RF cycle with 20 steps, while the $0.1\tau_{\mathrm{coll}}$ limits the per step collision probability to 10%. The acceleration $\Delta t_{\mathrm{acc}}$ and Courant–Friedrichs–Lewy (CFL) $\Delta t_{\mathrm{CFL}}$ limits restrict velocity changes and displacement, respectively; the latter limits displacement to half the smallest grid spacing. $\Delta t_{\mathrm{L}}$ controls Langevin relaxation updates, and $\Delta t_{\mathrm{rem}}$ is the time remaining until the next scheduled integration boundary, which prevents integration across the next PIC update, stage transition, or simulation endpoint. This scheme retains full three-dimensional particle dynamics while keeping the axisymmetric field solve computationally tractable. The key mathematical methods employed in the simulations are presented here; for further details regarding different parts, please refer to the Supporting Information Part II: Mathematical Methods.

**Coordinate system, static fields, and RF convention.**

Fields are represented on an axisymmetric $(r, z)$ grid, whereas ions retain three-dimensional Cartesian positions and velocities. Local radial field components are projected onto the Cartesian directions during particle integration, and collision updates retain all three velocity components. This formulation preserves three-dimensional particle motion while reducing field-storage and PIC computational costs.

The API with the S-lens of the custom-built ion trap instrument serves as a representative case, as it also provides a direct experimental benchmark for the simulations [29]. Its geometry and experimental design are shown in Supporting Information S1, and have been reported previously [3, 30]. The implementation is motivated by the coupled gas expansion, RF/DC focusing, electrode geometry, and ion current effects observed in S-lens interfaces [7, 9, 10]. Refined SIMION potential arrays are applied. The "field baking" step uses the project coordinating convention

$$X_{\mathrm{SIMION}} \to z, \qquad Y_{\mathrm{SIMION}} \to x, \qquad Z_{\mathrm{SIMION}} \to y \tag{2}$$

with the axisymmetric radial coordinate $r = \sqrt{x^2 + y^2}$ . For the present S-lens geometry, the precomputed electric field spans $z$ = 0–65 mm and $r$ = 0–20 mm with 10 μm grid spacing. The first electrode starts at $z$ = 5 mm, the capillary exit and source plane are placed at $z$ = 6 mm for the case simulations, as shown in Figure 2. Electric fields are obtained from the precomputed potentials using spline-based derivatives.

Two SIMION grids are processed, a DC potential $\phi_{\mathrm{DC}}(r, z)$ and a normalized RF basis potential $\phi_{\mathrm{RF,basis}}(r, z)$. The RF basis is generated with adjacent RF electrodes at $+1$ and $-1$ V; therefore, the runtime RF input is a peak voltage. For a particle at $(x, y, z)$ , the radial field component is projected into Cartesian space as

$$E_x = E_r \frac{x}{r}, \qquad E_y = E_r \frac{y}{r}, \qquad E_z = E_{\mathrm{axial}} \tag{3}$$

with zero transverse projection on the symmetry axis.

$$\mathbf{E}_{\mathrm{tot}}(\mathbf{r}, t) = \mathbf{E}_{\mathrm{DC}}(\mathbf{r}) + \mathbf{E}_{\mathrm{RF,basis}}(\mathbf{r}) \frac{V_{\mathrm{RF,peak}}}{V_{\mathrm{ref}}} \cos\,(2\pi f_{\mathrm{RF}} t + \phi_{\mathrm{RF}}) + \mathbf{E}_{\mathrm{SCE}}(\mathbf{r}, t) \tag{4}$$

where $V_{\mathrm{ref}} = 1$ V for the normalized RF basis and $\mathbf{E}_{\mathrm{SCE}}$ is the PIC space charge field.

**Gas field and electrode mask coupling.**

The gas field is imported from CFD simulation in the same global $(r, z)$ convention as the electric field processing. Runtime interpolation returns local axial and radial gas velocity, static temperature and pressure at every particle position. Gas to ion coupling is one-way in the present implementation. The prescribed CFD field affects particle transport, but ion motion does not update the neutral flow. This is appropriate for comparative ion optical studies in which the neutral solution is treated as a fixed operating condition [7, 9].

A separate electrode mask is generated from the SIMION geometry file and cached on the precomputed grid. During propagation, the nearest mask node and the precomputed surface distance field are queried for each active ion. A particle is classified as an electrode hit when it enters a metal cell or approaches the metal surface within the configured hit distance.

Coordinate consistency among SIMION potentials, gas flow fields (CFD setup details are described in Supporting Information Figure S3 and S4), and the electrode mask is checked after all sources are transformed into the common coordinate system (Figure 2). The diagnostic view spans the capillary exit and early S-lens region, where gas expansion, RF/DC focusing, and space charge first become strongly coupled. The production “field baking” uses a 10 µm static field grid over $z$ = 0–65 mm and $r$ = 0–20 mm. The displayed RF panel is scaled to 100 V. The shared capillary exit marker and metal contours provide a direct check of axial origin and mask alignment.

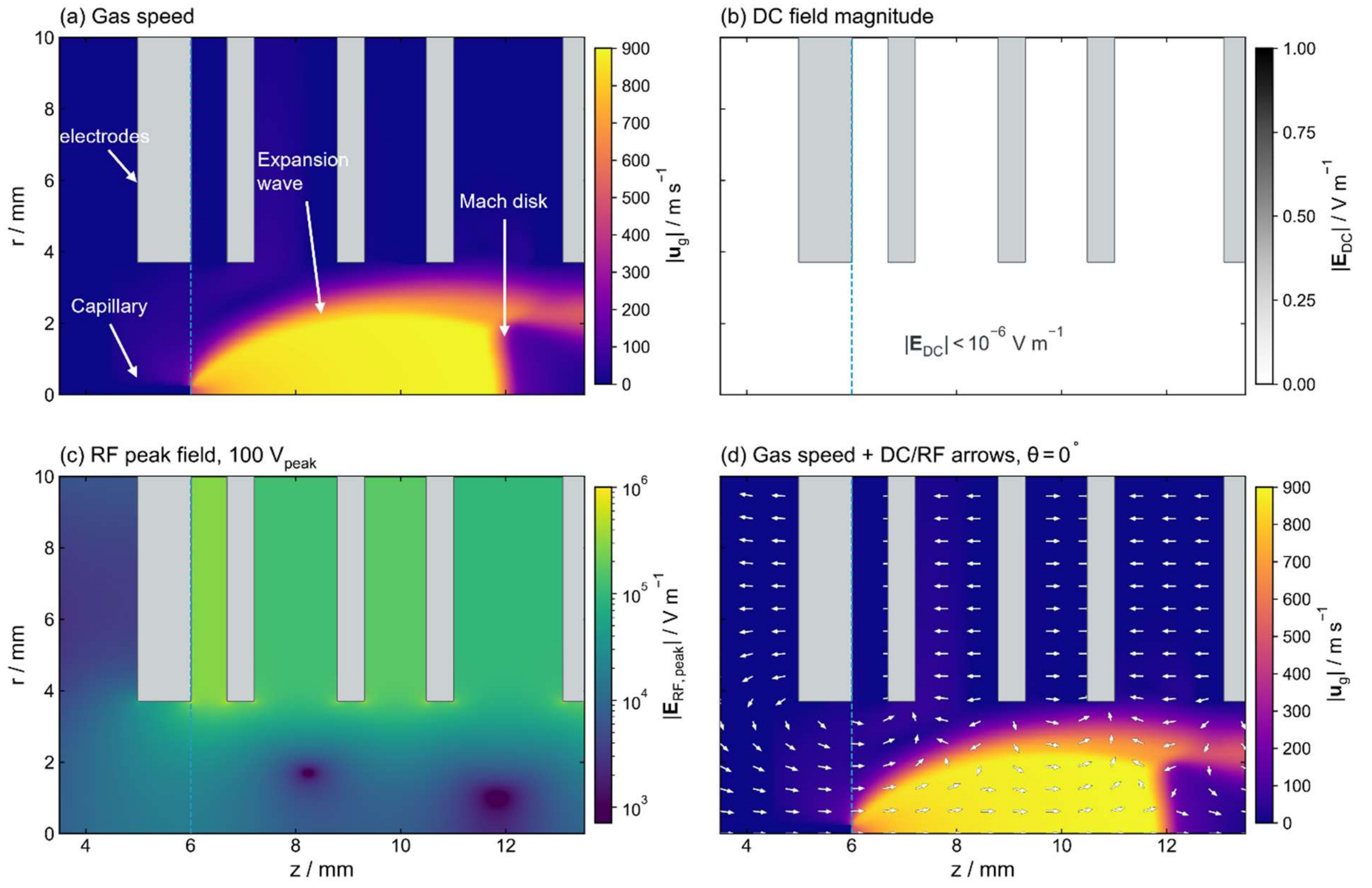

**Figure 2.** Coordinate alignment diagnostic near the custom API source entrance. (a) Gas flow magnitude, (b) DC field, (c) RF field, (d) total field direction. The blue line marks the capillary exit plane.

**Continuous current Gaussian source.**

The ion flux for the simulation is set by the ion current. Ions are continuously generated and grouped into weighted ion "packs" before implantation to simplify calculations. For current $I$, charge state $z_q$, and interval $\Delta t$, the source accumulator receives the corresponding number of represented real ions,

$$N_{\text{real}} = \frac{I\,\Delta t}{z_q e} \tag{5}$$

when the accumulator reaches the configured target ion pack weight $W$, one or more weighted ion packs are emitted. The prescribed current sets the total ionic charge injected per unit time, while the target ion pack weight controls the number of computational ion packs used to represent those ions. The emitted, active, and terminal represented weights are tracked independently for conservation checks.

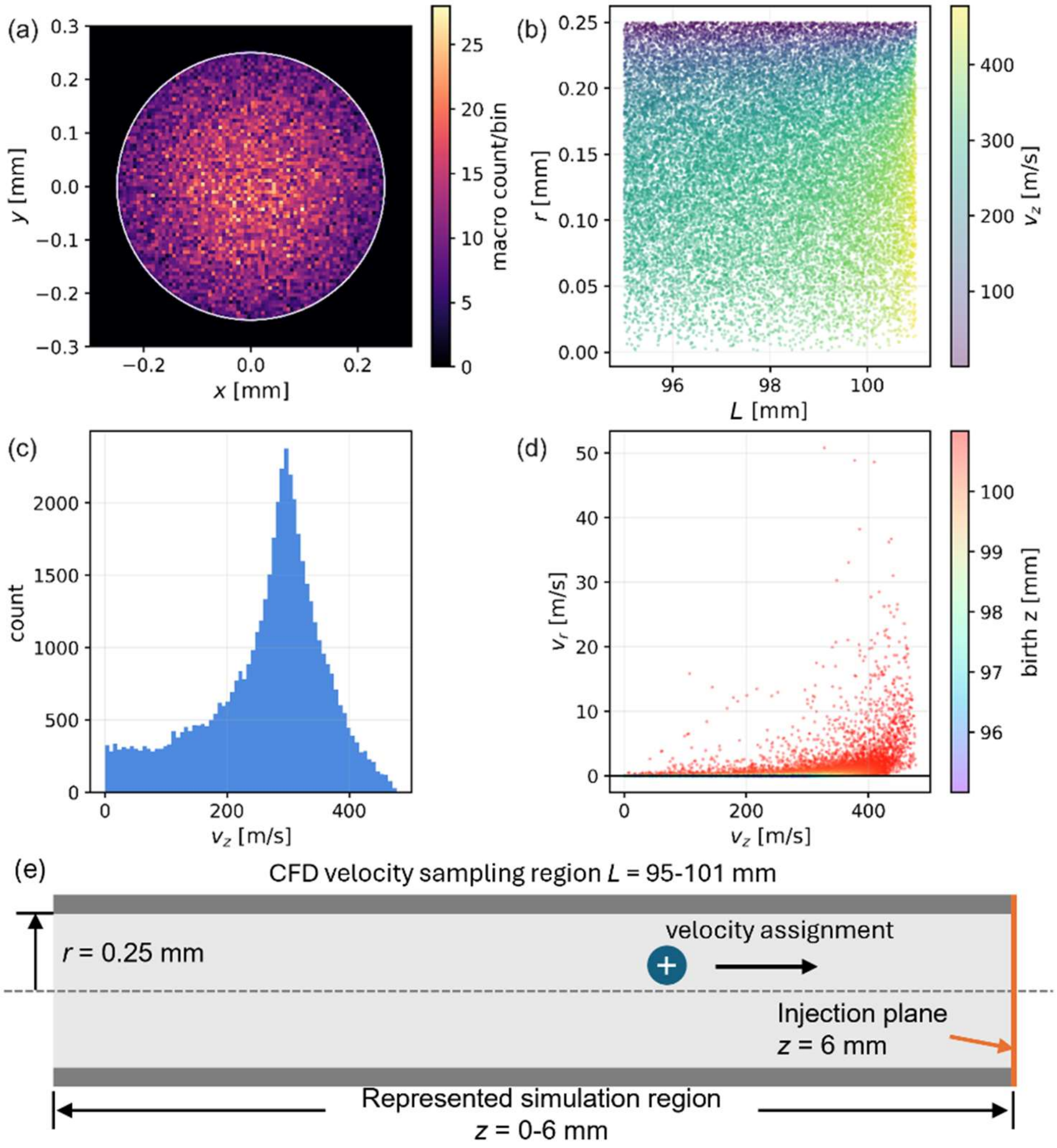

**Figure 3.** Continuous source sampling diagnostics. (a) Transverse source histogram. (b) Sampled capillary coordinates colored by axial gas velocity. (c) Initial axial velocity distribution. (d) Radial versus axial velocity, colored by capillary sampling position. (e) velocity-sampling region within the CFD-resolved capillary.

For each emitted weighted ion pack, transverse coordinates are sampled from a truncated two-dimensional Gaussian distribution with a standard deviation of $\sigma$ = 0.20 mm in each transverse Cartesian direction. Ion packs are loaded at a prescribed speed with radial boundary limitation of r = 0.25 mm which represents real capillary volume inside the source, this region is excluded from PIC charge deposition and electric field influence. They become active for transport upon reaching the capillary exit plane at z = 6 mm. Initial velocity is sampled from the measured CFD tube range at raw capillary coordinates $z$ = 95–101 mm, which maps to global $z$ = 0–6 mm. The local axial and radial gas velocities are interpolated from CFD result, and the radial component is projected onto the particle azimuth. This construction avoids prescribing a single cone velocity and retains the correlation between axial position and gas velocity near the capillary exit.

Figure 3(b) relates sampled capillary coordinates to the CFD axial velocity. Figure 3(c) shows the resulting axial velocity distribution, and Figure 3(d) shows the correlation between axial and radial velocity. Radial velocity becomes appreciable near the final portion of the 101 mm capillary, corresponding to the region that maps to the capillary exit in the precomputed field.

**Axisymmetric PIC space charge update.**

The PIC update calculates the electric field generated by the moving ion cloud. At each update, the current ion distribution determines the beam charge density, from which the space charge field is calculated. This field is added to the applied DC and RF fields for subsequent particle propagation.

Space charge effects are calculated using an axisymmetric PIC method with cloud-in-cell (CIC) charge deposition and field interpolation [31, 32]. Each active three-dimensional weighted ion pack is mapped onto $(r, z)$, and its weighted charge $qW$ is deposited to the four surrounding grid nodes using bilinear CIC weights. Because the grid represents a three-dimensional body of revolution, deposited charge is normalized by the corresponding annular control volume to obtain $\rho(r, z)$ in $C/m^3$.

The framework supports an AMG-preconditioned [33] conjugate-gradient Poisson solver using a Ruge-Stüben multigrid hierarchy [33, 34]. The cylindrical Poisson equation is discretized in finite-volume form over annular control volumes, yielding a sparse linear system. A multigrid hierarchy is constructed for the fixed grid and electrode mask and reused across PIC updates. The previous potential solution can be used as the initial guess. Convergence requires successful solver termination and an independently evaluated residual satisfying the specified tolerance. The iteration limit and tolerance are configurable; reaching the iteration limit without satisfying the convergence checks results in rejection of the solution. The accepted potential is differentiated to obtain the space charge field, which is interpolated to particle positions.

Production runs used a 201 × 651 PIC mesh with AMG preconditioning for conjugate-gradient iteration, a relative residual tolerance of $10^{-6}$, and a maximum of 150 iterations per solve. The multigrid hierarchy was constructed once, and warm starts were enabled for subsequent PIC updates.

**Particle transport, collisions, and effective internal temperature.**

Particle motion is integrated in three dimensions with an RK4 pusher under the instantaneous total field. At each micro step, the local gas field is sampled at the ion position, and the ion neutral collision rate is estimated from

$$\lambda = n_g \sigma_{\mathrm{ig}} \langle g \rangle \tag{6}$$

$$\langle g \rangle = \int_{\mathbb{R}^3} \left| \mathbf{v}_i - \mathbf{u}_g - \mathbf{c} \right| f_{\mathrm{M}}\left(\mathbf{c}; T_g\right) \mathrm{d}^3 c \tag{7}$$

where $n_g$is the neutral number density, $\sigma_{ig}$ is the configured collision cross section, $\mathbf{v}_i$ is the ion velocity, and $\mathbf{u}_g$ is the local gas velocity. The neutral thermal velocity $\mathbf{c}$ follows the normalized Maxwell–Boltzmann distribution $f_{\mathrm{M}}$ at the local gas temperature $T_g$. At zero drift, $\langle g \rangle = \sqrt{8k_B T_g/(\pi m_g)}$, where is the mass of one neutral molecule. The expression and its numerical evaluation are given in the Supporting Information Part II: Mathemathical Methods.

For each explicit collision, MuCITE applies IICT update to adjust the ion's translational velocity and internal energy. Based on a defined heat capacity model, the updated internal energy is converted into an effective internal temperature $T_{\mathrm{eff}}$. This approach accounts for the coupled transfer of momentum and internal energy, encompassing both collision-induced heating and cooling effects [24].

In the simulations presented below, an IICT-lite backend, derived from the IICT, was implemented, incorporating specific heat capacity profiles for Rhodamine B and protonated leucine enkephalin, respectively. IonSPA provides these heat capacity profiles, a state-dependent pseudo-atomic mass model, and dissociation rate functions derived from specified activation enthalpies and entropies [22, 23]. Furthermore, the system supports integration with IonSPA. A comparison of the two simulations can be found in Supporting Information S2.

In regions with high collision frequencies, explicit event sampling may dominate runtime performance due to the high frequency of collisions. The optional hybrid Langevin branch [35] therefore replaces selected explicit events in a prescribed axial window with aggregated velocity relaxation toward the local gas velocity plus a temperature dependent stochastic component. For the study cases, the window is $z$ = 6–7 mm. Its use for activation metrics is evaluated against matched explicit cases.

Dissociation is modeled as a temperature dependent loss of precursor ions in the pack. At every ion pack micro step, a transition state theory rate is evaluated from the current $T_{\text{eff}}$, $\Delta H^{\ddagger}$ and $\Delta S^{\ddagger}$ giving

$$SY(t) = \exp\left[-\int_0^t k\left(T_{\text{eff}}(\tau)\right) d\tau\right]. \quad (8)$$

The number of precursor ions in pack is reduced accordingly, and the weight is recorded as dissociated without generating separate product ion trajectories. The quantity is therefore reported as a predicted dissociated precursor survival yield $SY$.

**Terminal events, snapshots, and reported observables.**

Particles are classified as transmitted when they cross the detector plane within the detector radius. Other terminal classes are electrode hit, radial loss, and axial domain exit. For continuous sources, exit current $I_{\text{exit}}$ in a time bin $t_i$ is calculated from the sum of represented weights $W_j$ with charge $z_q e$ crossing the plane,

$$I_{\text{exit}}(t_i) = \frac{z_q e}{\Delta t} \sum_{j \in Z_{\text{exit}}(t_i)} W_j. \quad (9)$$

Exit beam size is reported as the weighted beam radius containing 95% of the transmitted represented weight; kinetic energy and $T_{\text{eff}}$ are summarized by the weighted median and the central 90% range (5th–95th percentiles).

## EXPERIMENTS

Ion transmission measurements were performed on a custom-built ion trap mass spectrometer equipped with an ESI source, which equipped with a S-lens and exit lens as ion guides [29]. Rhodamine B (100 μM) was sprayed in positive ion mode from a 1:1 (v/v) water/MeOH solution containing 2% acetic acid. The solution was introduced through a fused silica capillary at a flow rate of 1.5 μL min$^{-1}$ and a spray voltage of 2.4 kV. Ions were transferred into the first vacuum stage through a heated capillary (120 °C, 150 V) and were subsequently focused by the S-lens ion guide [3, 30]. The S-lens was followed by an exit electrode with a 2 mm aperture. The S-lens DC offset and the exit electrode were held at 130 V, giving no net DC potential difference between them. The pressure in this stage was maintained at 0.96 mbar. Ion currents were measured using a Faraday cup connected to a Keithley 6485 picoammeter, positioned 2 mm downstream of the exit lens.

An independent activation dataset was obtained using a Q-Exactive HF (QE HF) mass spectrometer (Thermo Fisher Scientific). The S-lens/exit lens assembly was operated in positive ion mode at 1.93 mbar. Protonated leucine enkephalin was generated from a 1 μM solution in 1:1 (v/v) acetonitrile/water containing 0.1% acetic acid and detected at *m/z* = 556.27 as the singly charged $[M+H]^+$ ion. The capillary temperature to 200 °C, and each spectrum was acquired for 30 s.

## RESULTS AND DISCUSSION

### Subsystem verification.

Pure field particle transport was benchmarked against a matched SIMION trajectory log and is documented further in the Supporting Information S3. The strict SIMION benchmark uses $z_{exit}$ = 65 mm. For 1000 identical initial rays, SIMION classified 989 ions as outside the work bench and 11 as electrode hits. MuCITE reproduced the same 989/11 classification and the same hit ion IDs. For 95% of the matched trajectories, the absolute errors were below 2.10 ns for time-of-flight and 8.92 meV for kinetic energy. This test verifies coordinate mapping, RF convention, field interpolation, electrode classification, and the RK4 trajectory path under the matched pure field conditions.

### Case study I: coupled transport in the custom S-lens interface.

The capillary exit region of an API source is typically the most dynamically complex part of the ion transfer process. Pressure driven flow through a capillary can produce a choked, supersonic jet at the exit. The associated gas acceleration, Mach disk formation, and adiabatic temperature variation produce a strongly nonuniform flow field. At the same time, ions are concentrated close to the capillary exit plane, where space charge effects are expected to be most pronounced.

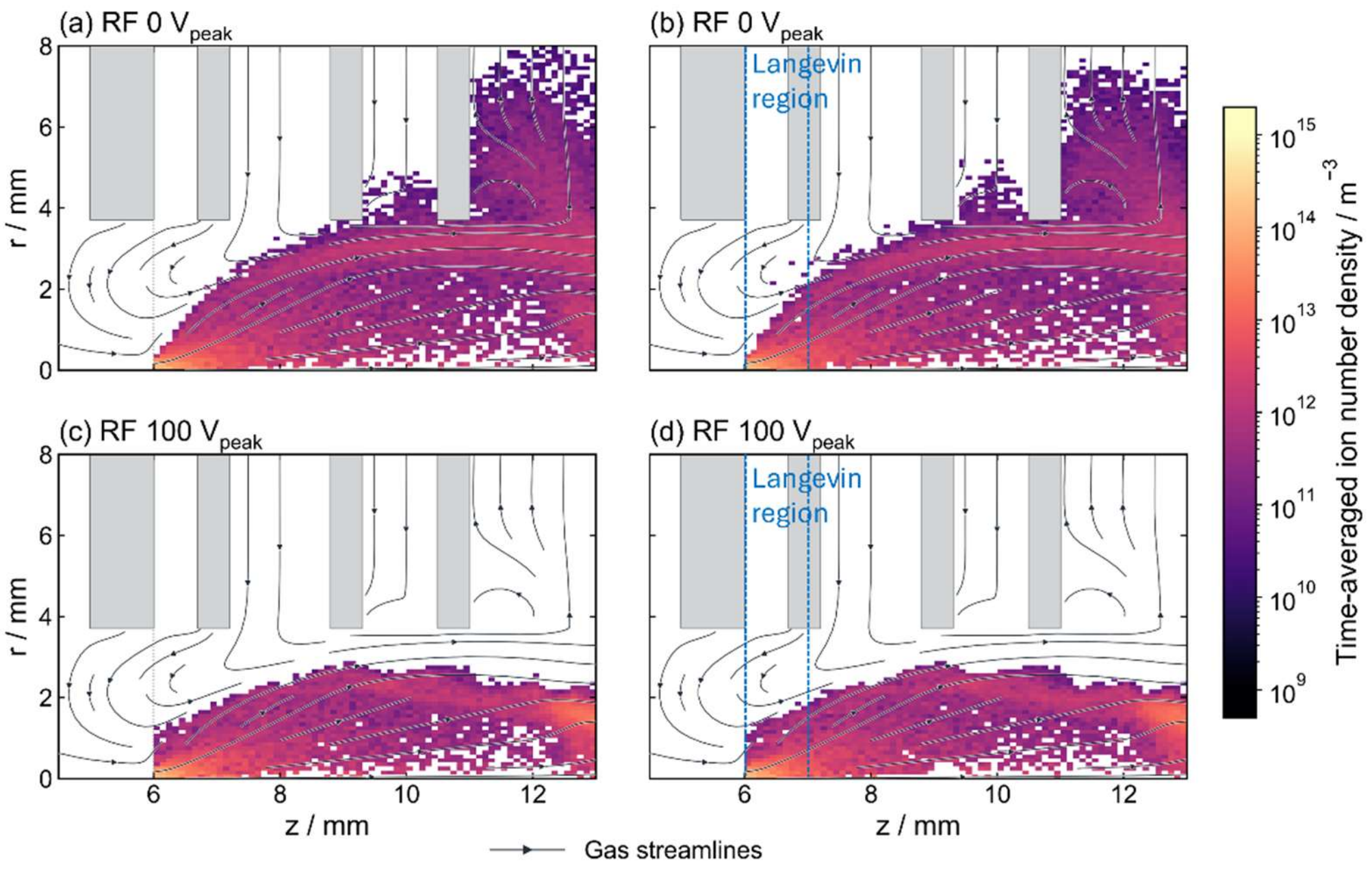

**Figure 4.** (a) Explicit collision and (b) hybrid Langevin simulation with RF 0 V at 2 nA. (c) Explicit collision and (d) hybrid Langevin simulation at 2 nA under RF 100 V. Grey blocks are electrodes. Blue lines in (b), (d) cover region using Langevin method.

The local ion distribution was examined at an injected current of 2 nA, RF amplitude of 100 V, pressure of 1 mbar (see Supporting Information S4 for parameters details). The simulation exit plane is $z_{exit}$ = 61 mm, 2 mm downstream of the exit lens. The hybrid Langevin branch was restricted to $z$ = 6–7 mm. With the RF off, as shown in Figure 4(a) and (b), the rapidly expanding gas jet initially entrains the ions downstream. In the absence of RF confinement, however, the radial component of the gas expansion drives most ions toward the electrode surfaces before effective axial transmission through the S-lens can be established. The hybrid calculation produces a slightly broader ion distribution within the Langevin region, whereas beyond this region it reproduces the downstream motion and dominant loss pattern obtained with the explicit model. Approximately 99% of the represented ion weight terminates through electrode collisions in both RF off simulations.

**Table 1.** Matched collision model checks at 2 nA.

| Collision model | Exit current / nA | Transmission | Median KE / eV | Median $T_{\mathrm{eff}}$ / K | Survival yield | Runtime / hrs |
|---|---|---|---|---|---|---|
| Explicit RF 0 V | 0.03 | 1.6% | 0.111 | 298 | 100% | 54.5 |
| Hybrid RF 0 V | 0.03 | 1.4% | 0.107 | 294 | 100% | 3.0 |
| Explicit RF 100 V | 1.50 | 74.9% | 0.099 | 304 | 59.8% | 56.5 |
| Hybrid RF 100 V | 1.49 | 74.7% | 0.101 | 304 | 56.7% | 2.6 |

When an RF amplitude of 100 V is applied (Figure 4 (c, d)), the ions initially follow the gas expansion at the capillary exit but are subsequently reshaped and confined by the combined effects of RF focusing and collisional momentum transfer. A region of inward radial gas velocity is visible near $z$ = 9–11 mm and r ≈ 2 mm, indicating a local radial flow reversal that also influences the ion trajectories.

Time resolved ion transfer process is considered in the simulation, as discussed in Supporting Information S5. Quantitatively, the explicit and hybrid calculations at 100 V differ in transmission by only 0.2% points and yield similar median exit kinetic energies and median effective internal temperatures, $T_{\mathrm{eff}}$, as summarized in Table 1. The hybrid calculation reduced elapsed runtime by a factor of 20 under these execution conditions. This speedup arises primarily from the Langevin treatment of the dense $z$ = 6–7 mm region, where the pressure decreases from around 150 to 30 mbar and explicit treatment would require many individual collision calculations. The hybrid approach is therefore suitable for large parameter sweeps and rapid design evaluation, provided that the Langevin region and switching conditions are first validated against representative explicit collision simulations.

**Current dependent transmission and electrode loss.**

The current sweep covers 0.5, 1, 2, 3, 4, and 6 nA at 80 V, 1 mbar and CCS of Rhodamine B in $N_2$ of $2.07 \times 10^{-18}$ $m^2$ [36]. Stable ion transmission decreases monotonically from 98.3% at 0.5 nA to 27.6% at 6 nA (Table 2). The corresponding increase in downstream electrode interception identifies current dependent space charge expansion as the dominant loss mechanism in this case.

The transmitted beam broadens with injected current. At $z$ = 61 mm, the exit beam radius (see Supporting Figure S8) containing 95% of the transmitted ion current increases from 1.63 mm at 0.5 nA to 2.39 mm at 6 nA. Changes in kinetic energy and survival yield reflect a radial redistribution of ions driven by space charge effects. This redistribution alters the RF-driven collisional activation and influences the filtering of the ion population by the exit aperture. At high ion currents, space charge effects cause the ion population to shift toward the outer regions, resulting in a reduction in axial kinetic energy. Simultaneously, this radial expansion modifies the RF field and collisional conditions encountered during ion sampling, which may lead to enhanced internal activation.

**Table 2.** Current sweep at 80 V RF amplitude.

| Input current / nA | Stable exit current / nA | Transmission | Radius enclosing 95% / mm | Median KE / eV | Median $T_{\mathrm{eff}}$ / K | Survival yield |
|---|---|---|---|---|---|---|
| 0.5 | 0.49 | 98.3% | 1.63 | 0.121 | 298 | 97.2% |
| 1.0 | 0.90 | 90.3% | 1.99 | 0.107 | 300 | 97.0% |
| 2.0 | 1.33 | 66.3% | 2.28 | 0.097 | 301 | 96.4% |
| 3.0 | 1.50 | 49.9% | 2.30 | 0.092 | 303 | 95.7% |
| 4.0 | 1.57 | 39.2% | 2.37 | 0.094 | 304 | 95.0% |
| 6.0 | 1.66 | 27.6% | 2.39 | 0.092 | 303 | 93.1% |

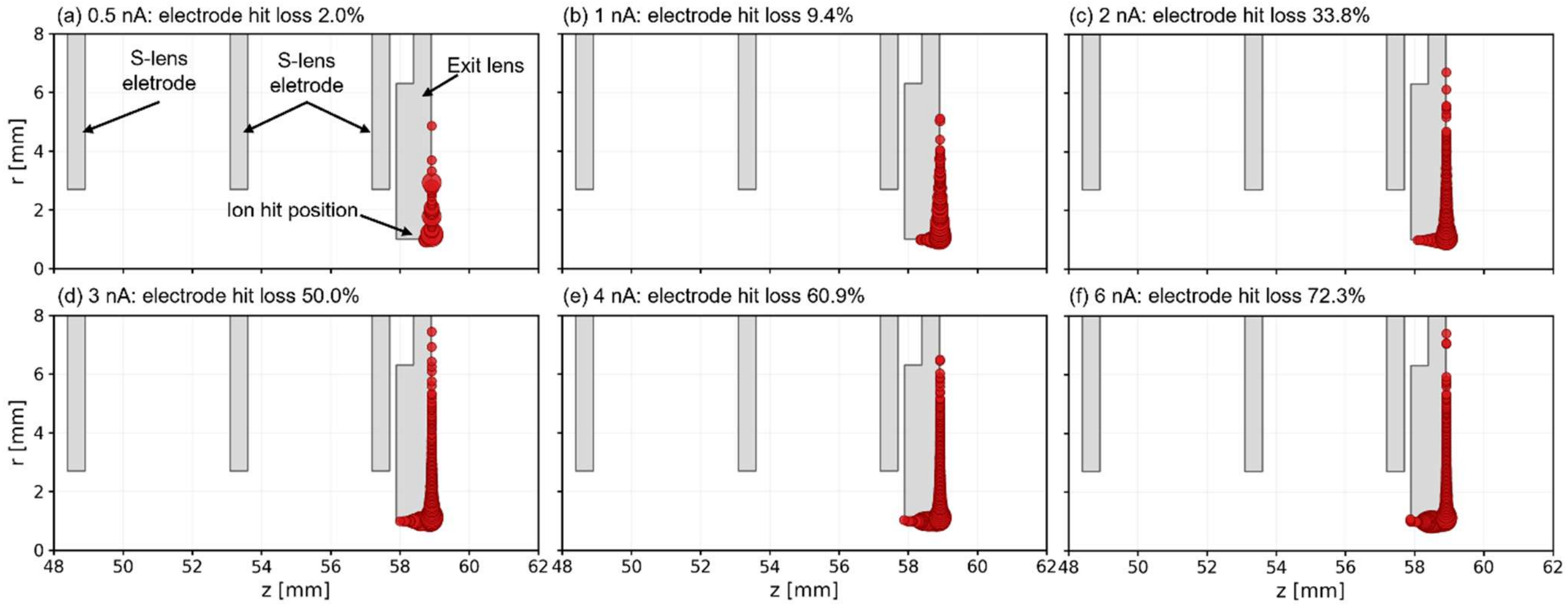

**Figure 5.** Electrode hit locations for injected currents of (a) 0.5, (b) 1, (c) 2, (d) 3, (e) 4, and (f) 6 nA. Gray regions show the electrode mask; red marker area is proportional to representing lost weight, larger marker means more ion losses.

The electrode loss fraction increases from 2.0% at 0.5 nA to 72.3% at 6 nA. Losses are concentrated near the exit lens aperture (Figure 5). The localization indicates that the 2 mm diameter, 1 mm thick aperture is the principal acceptance bottleneck under high current space charge expansion and can be therefore a direct geometry target for design modification.

**RF confinement, activation, and comparison with experiment.**

The simulation reproduces the main rising edge of the measured Rhodamine B transmission curve (Figure 6(a)). Simulated transmission increases from 1.3% at 0 V to 11.6% at 20 V, 47.8% at 50 V, and 88.0% at 160 V. The rising curve demonstrates progressively stronger radial confinement of the gas expanded plume. At 50 V, the simulated transmission of 47.8% agrees closely with the normalized experimental transmission of ~ 47.5%. At higher RF amplitudes, the simulated geometric transmission continues to increase, whereas the measured total ion signal reaches a plateau. RF induced dissociation above 80 V, as indicated in Figure 6(a) precursor ion current change, may partly explain this discrepancy. At high RF amplitudes, CID can change the mass-to-charge ratio, collision cross section, mobility, residence time, and spatial distribution of transmitted ions. Dissociation does not intrinsically increase total charge, but it can redistribute the local charge density and thereby modify the space charge field and transmission.

Other possible reasons arise from assembly induced misalignment or downstream flow and electrode effects in the experimental instrument, which can reduce effective ion optical acceptance and create localized transmission bottlenecks absent from the idealized axisymmetric model. Incomplete desolvation and space charge limitations on ion number may also contribute; because the measured current is not species selective, it may include residual clusters, precursor ions, and charged fragments.

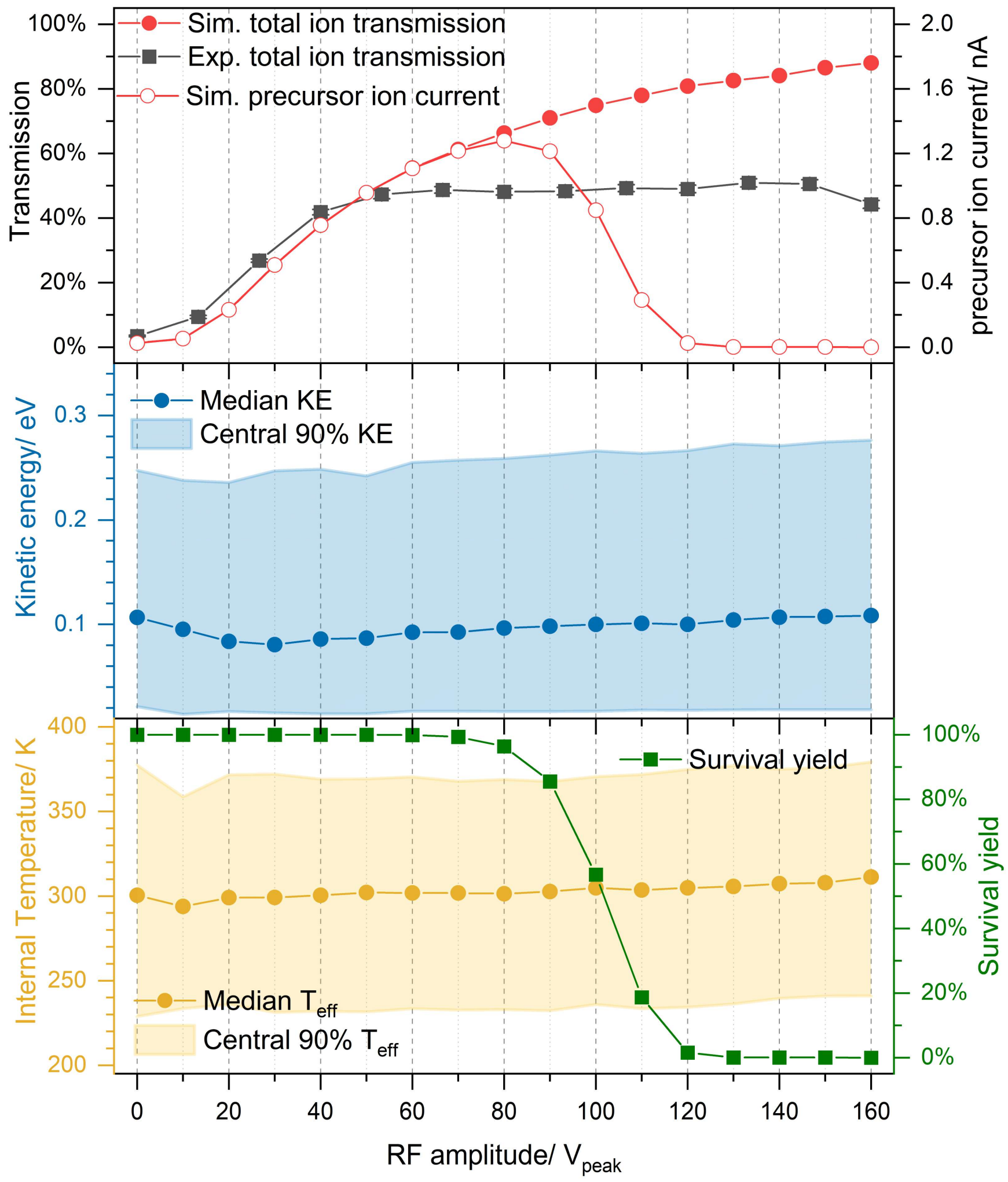

**Figure 6.** RF amplitude response for Rhodamine B. (a) Stable simulated transmission compared with experimental transmission and simulated precursor ion current. (b) Median and central 90% range (5th–95th percentiles) of kinetic energy. (c) $T_{\mathrm{eff}}$ and predicted survival yield.

The median exit kinetic energy varies only slightly, remaining within the range of 0.08–0.11 eV, while the median effective internal energy temperature stays around 300 K. These low exit state values do not imply that the transformation processes occurring throughout the ion source are entirely "gentle". Instead, ions experienced activation followed by collisional cooling.

Therefore, when ions exit the ion source they still exhibit low exit energies, even though they have undergone significant dissociation. The predicted precursor $SY$ remained 99.3% at RF amplitudes up to 70 V, which is consistent previous observations that varying the instrument RF level from 15% to 50% had no significant effect on in-source fragmentation [37], and was still 96.4% at 80 V, but fell sharply to 1.5% at 120 V. The simultaneous rapid loss of precursor survival above 80 V indicates a strongly nonlinear activation response to RF induced internal heating. The simulation demonstrates that precursor survival becomes highly sensitive to RF amplitude once the RF heating threshold is approached.

The influence of RF amplitudes illustrates the intended design use of MuCITE. Low RF amplitude limits sensitivity through insufficient confinement; increasing RF improves confinement and total transmission but eventually reduces precursor survival which no longer represent soft transfer. The operating point is therefore selected from a multi objective trade off rather than just geometric transmission alone.

**Case study II: leucine enkephalin survival yields.**

To evaluate the ability of MuCITE to predict collision induced ion activation trends, in-source fragmentation measurements of protonated leucine enkephalin on a QE HF ion source were used for experimental comparison. Two instrument settings were varied independently: the S-lens RF level and the in-source CID energy. The S-lens level is an instrument-defined control factor used with the selected scan mass range to determine the RF amplitude. The in-source CID energy, reported in eV, controls source activation. The S-lens level varied from 0 to 100 and the in-source CID energy was scanned from 0 to 100 eV. Figure 7(a) presents representative breakdown curves obtained at an S-lens level of 60.

At low collision energies, the protonated precursor ion $[M+H]^+$ was dominant. As the energy increased, the precursor signal decreased markedly above 40 eV, while product ions including $b_4^+$, $a_4^+$, $y_2^+$ and $b_3^+$ gradually appeared. Some products reached their maximum intensities at intermediate energies and subsequently decreased, suggesting further secondary dissociation, whereas other high energy products continued to increase. Figure 7(a) reflects not only precursor depletion but also redistribution among competing and sequential fragmentation pathways.

The experimental survival yield was defined as the intensities of the precursor relative to the combined the summed integrated intensities of the selected fragment ions and precursor ion:

$$SY = \frac{I_{MH^+}}{I_{MH^+} + \sum I_{F^+}} \times 100\% \tag{7}$$

Simulations used precomputed electric and gas flow fields for the QE HF S-lens geometry with chamber pressure of 2 mbar and capillary wall temperature of 473 K. Protonated leucine enkephalin was represented by $m/z$ = 556.27, a charge state of +1, a CCS in $N_2$ of $2.30 \times 10^{-18}$ $m^2$ [38], and 78 atoms. Conversion between ion internal energy and effective internal temperature employed the peptide heat-capacity profile. The activation enthalpy and entropy for the dissociation kinetics were $\Delta H^\ddagger = 111.7\ \mathrm{kJ\ mol^{-1}}$ and $\Delta S^\ddagger = -38.1\ \mathrm{J\ mol^{-1}\ K^{-1}}$, respectively [39]. The peptide heat-capacity profile provided by IonSPA is applied for the IICT model, which is used to convert internal energy into effective internal temperature. Note, however, that the collision energy transfer parameterization is based on molecular dynamics simulations of globular protein ions [25]. Applying this parameterization to leucine enkephalin therefore may introduce uncertainty in the predicted internal temperatures. The effective temperatures and dissociation fractions presented in this case study are intended to assess only the activation trends associated with the model.

The instrument's in-source CID energy is a nominal source-activation setting reported in eV and cannot be directly equated with a single DC potential difference. In our simulations, activation depends on both the applied DC bias and RF-induced heating. We therefore examined combinations of DC bias and RF amplitude to identify conditions that approximate the experimental activation response. Precomputed DC fields varied the S-lens to exit-lens potential difference from 0 to 100 V, with the exit lens held at 0 V. This DC scan was performed at RF peak amplitudes of 30, 50, 70, 80, 90 and 100 V. The hybrid Langevin mode is applied in this section. As noted earlier, there is virtually no difference between the explicit and hybrid modes. We provide comparative results in the Supporting Information S7.

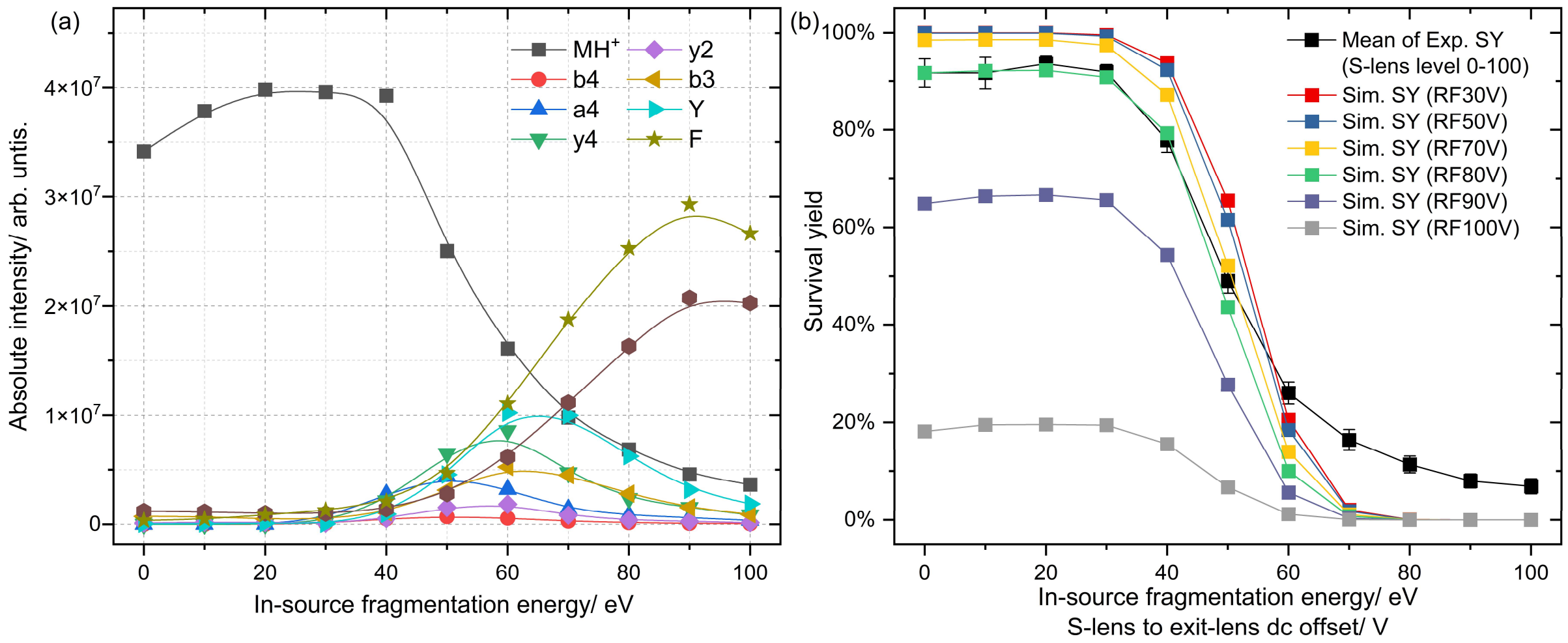

**Figure 7.** (a) Experimental breakdown curves over the in-source fragmentation energy scan under S-lens level 60. (b) Experimental mean dissociated precursor *SY* of S-lens level 0–100, compared with the simulated RF amplitude 30, 50, 70, 80, 90 and 100 V dissociated precursor *SY* for DC biases of 0–100 V.

Experimental and simulated results both reveal similar low to high activation transitions (Figure 7(b)). When RF amplitude is below 80 V, the SY trends reproduced the dissociation baseline that began at DC offset of 40 V. At high DC offsets, all simulated curves converged towards complete dissociation. As the RF amplitude is increased to 90 V and 100 V, the dissociation effect induced by RF heating begins to emerge. An increase of only 10% in high RF amplitudes can substantially enhance ion activation. Because the dissociation rate depends exponentially on internal temperature, small differences in RF energy near the dissociation threshold can produce large changes in the fragmentation fraction.

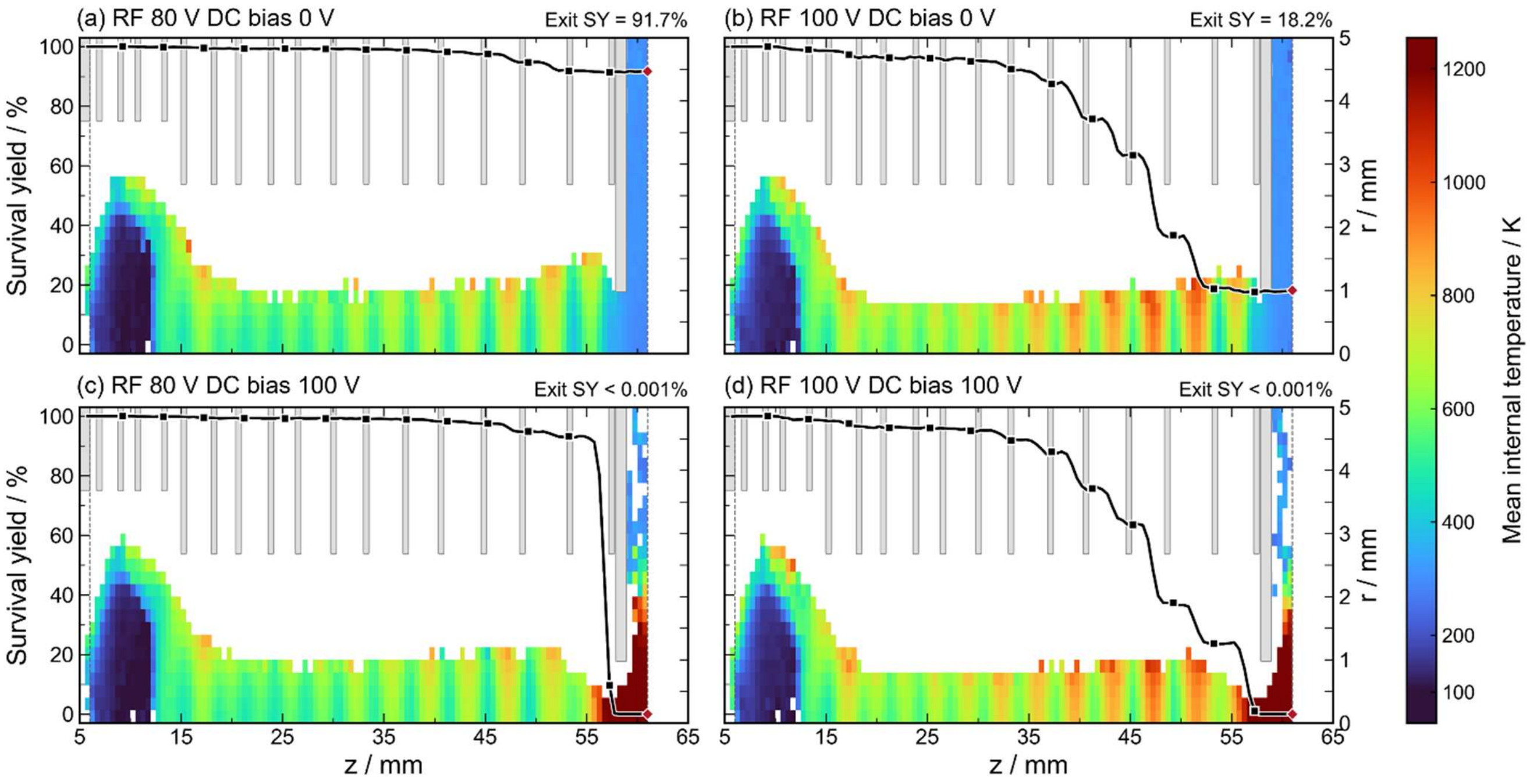

**Figure 8.** Spatially resolved internal temperature and precursor survival of protonated leucine enkephalin at RF amplitudes of 80 V with S-lens to exit lens DC offsets of 0 V(a) and 100 V(c) and RF of 100 V with DC biases of 0 V (b) and 100 V (d). Colors show the ions mean internal temperature. Black curves indicate precursor ion survival yields (left axis).

Figure 8 shows the mean internal temperature when ions flying though the position and the precursor survival yield changes along S-lens. A low-temperature region emerges near the capillary exit, resulting from the collisional cooling of ions by the supersonic expansion gas. Further downstream, the internal energy temperature rises and exhibits radial variation.

With an RF amplitude of 80 V and DC bias of 0 V (Figure 8(a)), the SY line shows slightly decline along the ion transport path, reaching 91.7% at the exit. Although there are some regions with higher local internal temperatures (over 800 K), ions cool to near room temperature by the time they reach the exit. Upon applying a 100 V DC bias, a distinct high-temperature region appears near the exit, accompanied by a sharp drop in the local survival rate at z ≈ 56 – 59 mm (Figure 8(c)). The predicted SY at the exit falls to 0%, indicating substantial additional activation associated with downstream acceleration and ion–gas collisions.

When the RF amplitude is 100 V, the rise in internal temperature within the S-lens becomes more pronounced. Even in the absence of DC bias, the SY gradually drops substantially within the S-lens, falling to 18.2% by the exit (Figure 8(b)). This heating occurs between z ≈ 35 and 55 mm, where the gaps between the electrodes are larger. Stronger electric fields accelerate the ions more vigorously, causing fragmentation due to the elevated temperature, even within a short timeframe. Applying a 100 V DC bias triggers further activation near the exit, reducing the survival rate of the remaining precursors to 0% (Figure 8(d)). These results distinguish between the activation induced by the RF field during ion transport and the additional activation generated by the DC bias in the exit region. At the same time, this explains why a "low exit temperature" alone cannot guarantee a sufficiently "gentle" ion-transport process.

From the perspective of ion source optimization, transmission and activation exhibit different voltage dependences. This separation provides scope for multi-objective optimization: lower DC offsets can maintain high transmission with gentler ion transfer, whereas higher DC offsets or RF amplitudes can provide controlled in-source activation. MuCITE can thus be used to identify RF/DC parameter combinations that satisfy target transmission, internal temperature, or dissociation requirements.

## CONCLUSION

MuCITE provides a method for continuous ion transport simulation through atmospheric pressure interfaces in mass spectrometry. It combines externally calculated DC/RF and gas flow fields with three-dimensional weighted ion pack propagation, axisymmetric PIC space charge, IICT based collision and internal energy updates, and spatially resolved terminal event accounting. The pure field benchmark establishes particle level agreement with SIMION for the validated field path. The two S-lens interfaces show how current, RF amplitude, and DC bias jointly affect transmission, localized loss, exit beam width, kinetic energy, effective temperature, and precursor dissociation propensity.

In simulations of the custom-built API source, increasing current broadens the beam and concentrates losses at the exit aperture. Increasing RF improves confinement but also raises the activation diagnostic. In the Q-Exactive HF case, MuCITE shows ability to identify RF/DC parameter combinations that satisfy target transmission, internal temperature, or dissociation requirements. These results demonstrate how MuCITE can support comparative source design and parameter selection for sensitivity, soft transfer, and controlled in-source activation. Future work will extend the field and space charge calculations to fully three-dimensional geometries and support mixed ion populations with species specific masses, charge states, collision cross sections, and internal energy parameters.

## ACKNOWLEDGMENT

We thank Flaela Kalemi for sample preparation and Kevin Li for assistance with experiments on the ion trap setup. We thank the reviewers for pointing out blind spots in the previous collision and dissociation model. Their input has helped us significantly improved the quality of the results obtained by MuCITE compared to the initial code.

**Supplementary information**

Part I: Description of geometry and imported-field diagnostics, IICT backend validation, pure-field SIMION verification, production simulation parameters, time resolved ion transfer, current-dependent exit distributions and comparison of explicit and hybrid Langevin modes. Part II: Mathematical methods described the detailed mathematical methods used in the simulation framework. (PDF)

**Data availability**

The software is available at https://github.com/gabelicagroup/MuCITE

## REFERENCES

(1) Fenn, J. B.; Mann, M.; Meng, C. K.; Wong, S. F.; Whitehouse, C. M. Electrospray Ionization for Mass Spectrometry of Large Biomolecules. *Science* **1989**, *246* (4926), 64-71.

(2) Covey, T. Where have all the ions gone, long time passing? Tandem quadrupole mass spectrometers with atmospheric pressure ionization sensitivity gains since the mid 1970s. A perspective. *Rapid Communications in Mass Spectrometry* **2022**, *39* (S1), e9354.

(3) Yan, Y.; Schmitt, L.; Khramchenkova, A.; Lengyel, J. Ion transmission in an electrospray ionization mass spectrometry interface using an S-lens. *Journal of Mass Spectrometry* **2023**, *58* (7), e4955.

(4) Shaffer, S. A.; Tang, K.; Anderson, G. A.; Prior, D. C.; Udseth, H. R.; Smith, R. D. A Novel Ion Funnel for Focusing Ions at Elevated Pressure Using Electrospray Ionization Mass Spectrometry. *Rapid Communications in Mass Spectrometry* **1997**, *11* (16), 1813-1817.

(5) Shaffer, S. A.; Prior, D. C.; Anderson, G. A.; Udseth, H. R.; Smith, R. D. An Ion Funnel Interface for Improved Ion Focusing and Sensitivity Using Electrospray Ionization Mass Spectrometry. *Analytical Chemistry* **1998**, *70* (19), 4111-4119.

(6) Kelly, R. T.; Tolmachev, A. V.; Page, J. S.; Tang, K.; Smith, R. D. The Ion Funnel: Theory, Implementations, and Applications. *Mass Spectrometry Reviews* **2010**, *29* (2), 294-312.

(7) Wissdorf, W.; Lorenz, M.; Poehler, T.; Hoenen, H.; Benter, T. Atmospheric Pressure Ion Source Development: Experimental Validation of Simulated Ion Trajectories within Complex

Flow and Electrical Fields. *Journal of the American Society for Mass Spectrometry* **2013**, *24* (10), 1456-1466.

(8) Gimelshein, S.; Lilly, T.; Moskovets, E. Numerical Analysis of Ion Funnel Transmission Efficiency in an API MS System with a Continuum/Microscopic Approach. *Journal of the American Society for Mass Spectrometry* **2015**, *26* (11), 1911-1922.

(9) Zhou, X.; Ouyang, Z. Following the Ions through a Mass Spectrometer with Atmospheric Pressure Interface: Simulation of Complete Ion Trajectories from Ion Source to Mass Analyzer. *Analytical Chemistry* **2016**, *88* (14), 7033-7040.

(10) Song, J.; Zhang, D.; Dai, X.; Huang, Z.; Fang, X.; Tian, D.; Jiang, Y. Numerical Analysis and Quantification of Transfer Efficiency Coupled with Capillary and Quadrupole Ion Guide in an API MS System. *Journal of the American Society for Mass Spectrometry* **2024**, *35* (7), 1497-1506.

(11) Zhou, X. Numerical Simulation for Mass Spectrometry Instrumentation. *International Journal of Mass Spectrometry* **2020**, *458*, 116439.

(12) Ewing, S. A.; Donor, M. T.; Wilson, J. W.; Prell, J. S. Collidoscope: an improved tool for computing collisional cross-sections with the trajectory method. *Journal of The American Society for Mass Spectrometry* **2017**, *28* (4), 587-596.

(13) Garimella, S.; Zhou, X.; Ouyang, Z. Simulation of rarefied gas flows in atmospheric pressure interfaces for mass spectrometry systems. *Journal of the American Society for Mass Spectrometry* **2013**, *24* (12), 1890-1899.

(14) Dahl, D. A. SIMION for the Personal Computer in Reflection. *International Journal of Mass Spectrometry* **2000**, *200* (1-3), 3-25.

(15) Manura, D.; Dahl, D. A. *SIMION 8.2 User Manual*; 2020.

(16) Zimmerman, W. B. J. Multiphysics modeling with finite element methods. In *Multiphysics modeling with finite element methods*, Vol. 18; World Scientific Publishing Company, 2006.

(17) Giberson, C.; Singh, R. K.; Chun, J.; Huntley, A. P.; Zhong, J.; Ibrahim, Y. M.; Schenter, G. K.; Lee, J.-Y.; Garimella, S. V. SimELIT: A Novel GUI-Based Comprehensive Ion Trajectory Simulation Software for Mass Spectrometry. *Journal of the American Society for Mass Spectrometry* **2022**, *33* (8), 1453-1457.

(18) Rajkovic, M.; Benter, S.; Hammelrath, M.; Thinius, M.; Benter, T.; Wissdorf, W. IDSimF: An Open Source Framework for the Simulation of Molecular Ion Dynamics in Mass Spectrometry and Ion Mobility Spectrometry. *Journal of the American Society for Mass Spectrometry* **2024**, *35* (7), 1451-1460.

(19) Rajkovic, M.; Benter, T.; Wissdorf, W. Molecular Dynamics Based Modeling of Ion Neutral Collisions in an Open Ion Trajectory Simulation Framework. *Journal of the American Society for Mass Spectrometry* **2023**, *34* (10), 2156-2165.

(20) Wissdorf, W.; Thinius, M.; Benter, T. Simulation of space charge effects in Fourier transform quadrupole ion traps (FT-QITs). *Journal of the American Society for Mass Spectrometry* **2024**, *35* (12), 2969-2983.

(21) Schaefer, C.; Jašík, J.; Španěl, P. ICARION: A Modular Framework for Ion Trajectory Simulation in Electric Fields and Neutral Gas Environments. *Journal of the American Society for Mass Spectrometry* **2026**, *37* (10), 2483–2496.

(22) Paris, L. R.; Green, A. W.; Prell, J. S. Computed Vibrational Heat Capacities for Gas-Phase Biomolecular Ions. *Journal of the American Society for Mass Spectrometry* **2025**, *36* (4), 862-872.

(23) Shepherd, S. O.; Green, A. W.; Wylie, E. P.; Newton, K. R.; Kurulugama, R. T.; Prell, J. S. Determination of Thermochemical Barriers in Multiple-Collision Induced Dissociation Experiments on Gas-Phase Protein Complexes. *Journal of the American Chemical Society* **2025**, *147* (51), 46854-46870.

(24) Prell, J. S. Modeling collisional kinetic energy damping, heating, and cooling of ions in mass spectrometers: A tutorial perspective. *International Journal of Mass Spectrometry* **2024**, *504*, 117290.

(25) Green, A. W.; Prell, J. S. Modeling Energy Transfer Efficiency in Collision-Induced Dissociation and Unfolding Using Molecular Dynamics Simulations. *Journal of the American Society for Mass Spectrometry* **2026**, *37* (10), 2552-2564.

(26) Shepherd, S. O.; Green, A. W.; Wylie, E. P.; Newton, K. R.; Kurulugama, R. T.; Prell, J. S. Determination of Thermochemical Barriers in Multiple-Collision Induced Dissociation Experiments on Gas-Phase Protein Complexes. *Journal of the American Chemical Society* **2025**, *147* (51), 46854.

(27) Newton, K. R.; Green, A. W.; Prell, J. S. SimIonSPA: Collision Heating and Cooling in Mass Spectrometry Collision Cells with Simion. In *74th ASMS Conference on Mass Spectrometry and Allied Topics*, San Diego, CA, 2026; American Society for Mass Spectrometry: Vol. ThP 142

(28) Newton, K. R.; Green, A. W.; Roush, A. E.; Bush, M. F.; Prell, J. S. SimIonSPA: Collision-Resolved Modeling of Protein Ion Temperatures in a SLIM model system. In *73rd ASMS Conference on Mass Spectrometry and Allied Topics*, Baltimore, MD, 2025; Vol. TP 375

(29) Li, K.; Yan, Y.; Lengyel, J. Ion trap mass spectrometer for kinetic studies of subnanometer particles. *Aerosol Science and Technology* **2026**, *60* (10), 1145-1156.

(30) Yan, Y. Design of an Ion Beam Setup for Cluster Reactivity and Nucleation Studies. Doctoral dissertation, Technische Universität München, 2025.

(31) Hockney, R. W.; Eastwood, J. W. *Computer Simulation Using Particles*; Adam Hilger, 1988.

(32) Birdsall, C. K.; Langdon, A. B. *Plasma Physics via Computer Simulation*; McGraw-Hill, 1985.

(33) Bell, N.; Olson, L. N.; Schroder, J.; Southworth, B. PyAMG: Algebraic Multigrid Solvers in Python. *Journal of Open Source Software* **2023**, *8*, 5495.

(34) Ruge, J. W.; Stüben, K. Algebraic Multigrid. In *Multigrid Methods*, McCormick, S. F. Ed.; Frontiers in Applied Mathematics, Vol. 3; SIAM, 1987; pp 73-130.

(35) Ermak, D. L.; Buckholz, H. Numerical Integration of the Langevin Equation: Monte Carlo Simulation. *Journal of Computational Physics* **1980**, *35* (2), 169-182.

(36) Zhou, Z.; Luo, M.; Chen, X.; Yin, Y.; Xiong, X.; Wang, R.; Zhu, Z.-J. Ion mobility collision cross section atlas for known and unknown metabolite annotation in untargeted metabolomics. *Nature Communications* **2020**, *11* (1), 4334.

(37) Criscuolo, A.; Zeller, M.; Fedorova, M. Evaluation of lipid in-source fragmentation on different orbitrap-based mass spectrometers. *Journal of the American Society for Mass Spectrometry* **2020**, *31* (2), 463-466.

(38) Nichols, C. M.; May, J. C.; Sherrod, S. D.; McLean, J. A. Automated flow injection method for the high precision determination of drift tube ion mobility collision cross sections. *Analyst* **2018**, *143* (7), 1556-1559.

(39) Sztáray, J.; Memboeuf, A.; Drahos, L.; Vékey, K. Leucine enkephalin—a mass spectrometry standard. *Mass Spectrometry Reviews* **2011**, *30* (2), 298-320.

Supporting Information for

# Multiphysics Simulation Framework for Continuous Ion Transport and Energetics in Atmospheric-Pressure Interfaces

*Yihui Yan[1], Jozef Lengyel[2], Frédéric Rosu[1], and Valérie Gabelica[1*]*

[1]School of Pharmaceutical Sciences, University of Geneva, 1205 Geneva, Switzerland

[2]School of Natural Sciences, Technical University of Munich, 85748 Garching, Germany

* Corresponding Author: Valérie Gabelica( Valerie.Gabelica@unige.ch)

# Table of Contents

# Part I: Simulations Methods and Validation Details

## S1. Geometry and imported-field diagnostics

Production simulations were executed on a Dell Pro Max Tower 2 (FCT2250) with an Intel Core Ultra 9 285K CPU (24 cores), 128 GB DDR5 RAM, and an NVIDIA RTX Pro 5000 Blackwell GPU. The production environment used Python 3.11.9 and Taichi 1.7.4, with up to eight independent cases running in parallel. Each result directory retains the full command line and simulation summary used to reconstruct the case.

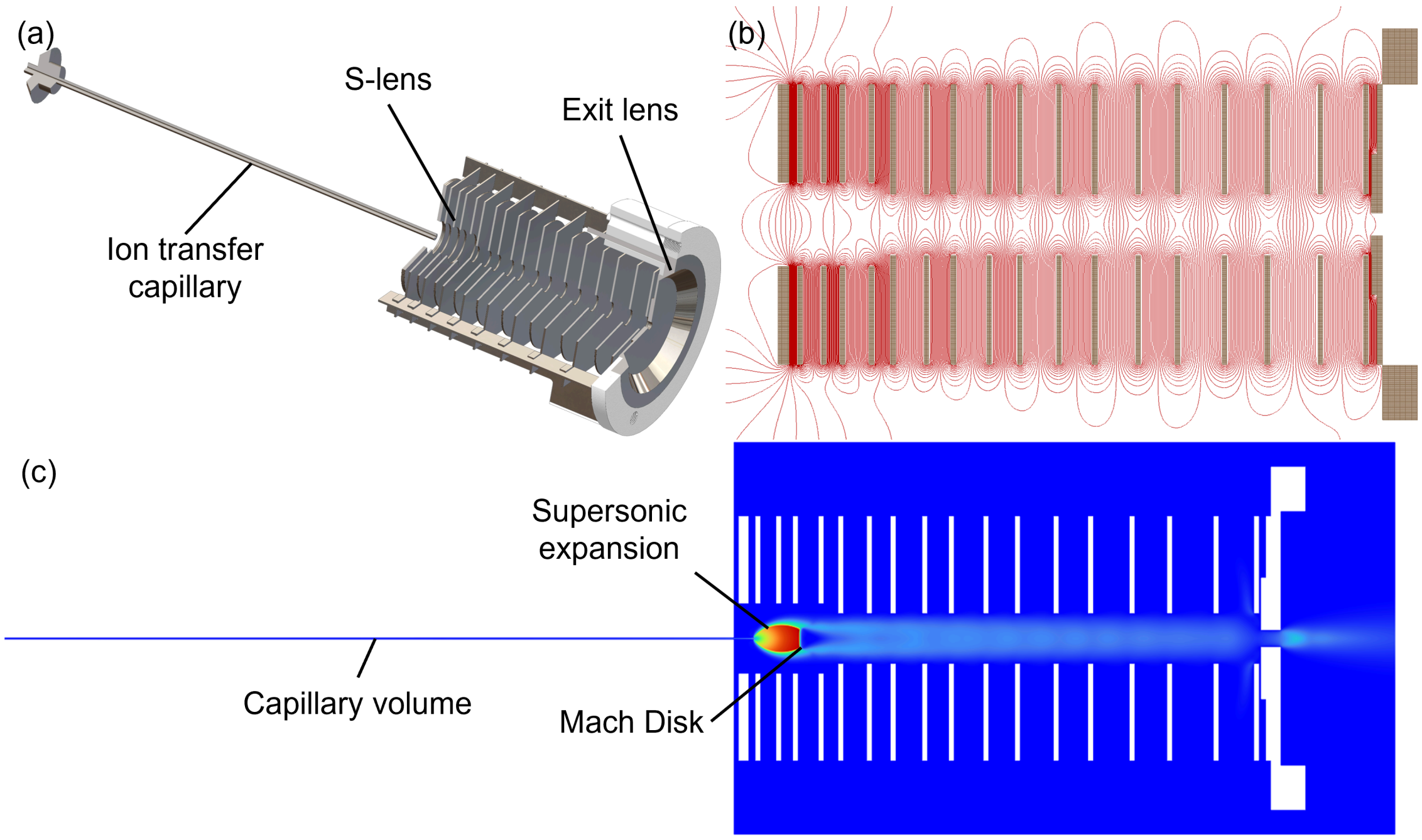

Figure S1. Geometry and fields of the modeled atmospheric pressure ionization interface. (a) Sectional view of ion-transfer capillary, S-lens, and exit lens assembly. (b) Combined DC and RF potential contours in the lens region, simulated with SIMION 2026. (c) CFD Mach-number distribution, based on long capillary and S-lens geometry, showing supersonic expansion and the downstream Mach disk on Ansys Fluent Student.

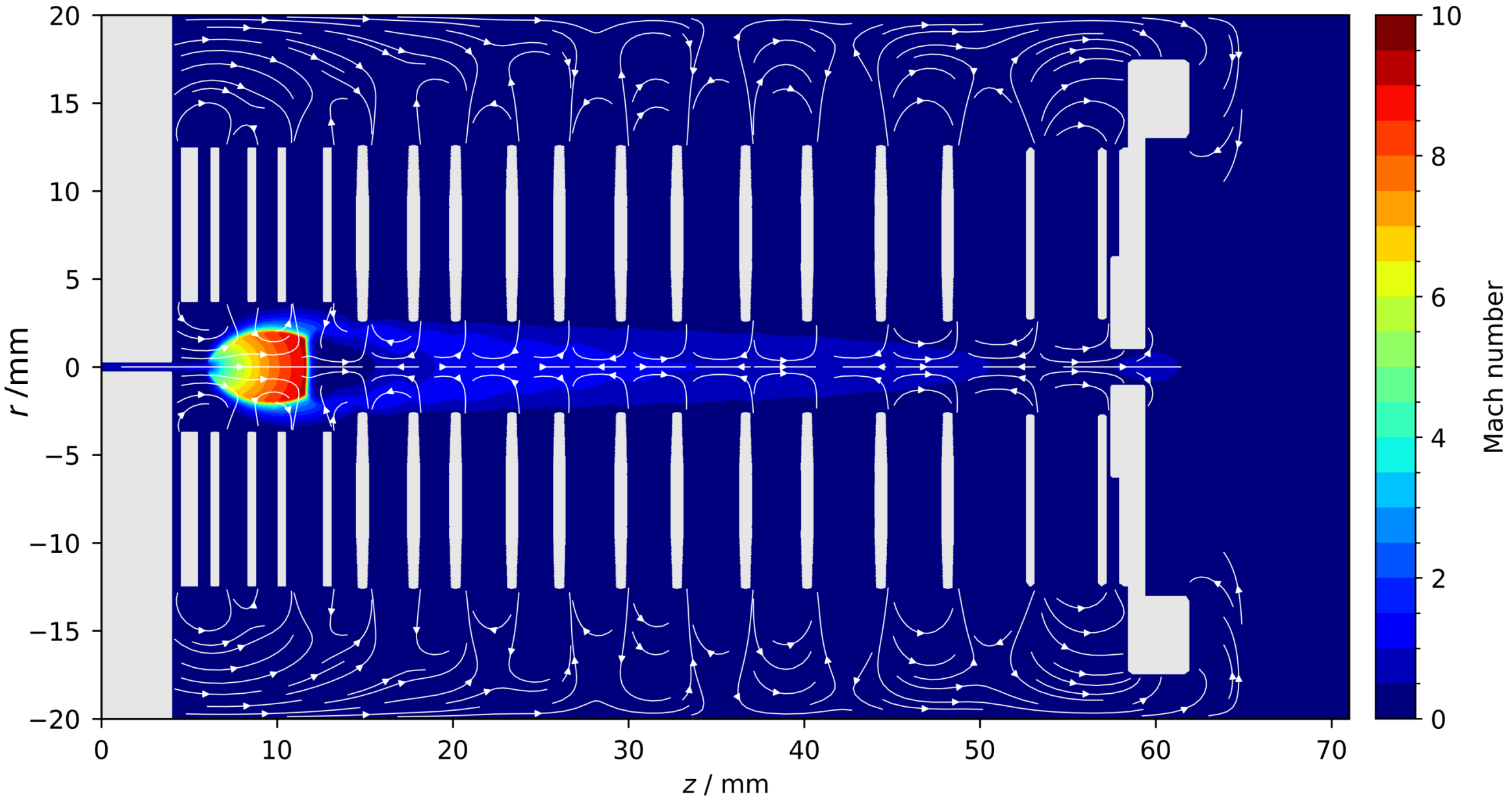

Figure S2. Simulation workplace is plotted with Mach-number distribution overlaid with electric-field lines at a DC offset of 50 V, an RF amplitude of 100 V peak to ground. White lines and arrows indicate the electric-field direction. The meridional section is mirrored about the axis, with the capillary exit at $z = 6$ mm. Grey regions lie outside the plotted fluid domain.

The CFD was simulated with Ansys Fluent Student. Gas is nitrogen represented as an ideal gas with the energy equation enabled. The initial gas temperature is 300 K. Capillary-wall temperatures are 400 K for the custom-API and 473 K for the Q-Exactive S-lens API. Turbulence is represented with the $k$–$\omega$ SST model. Momentum and energy use second-order upwind discretization; pressure uses a second-order pressure scheme. Reported residual convergence criteria are $10^{-4}$, and the exported solver record gives a final net mass-flow imbalance below $10^{-10}$ in its recorded units.

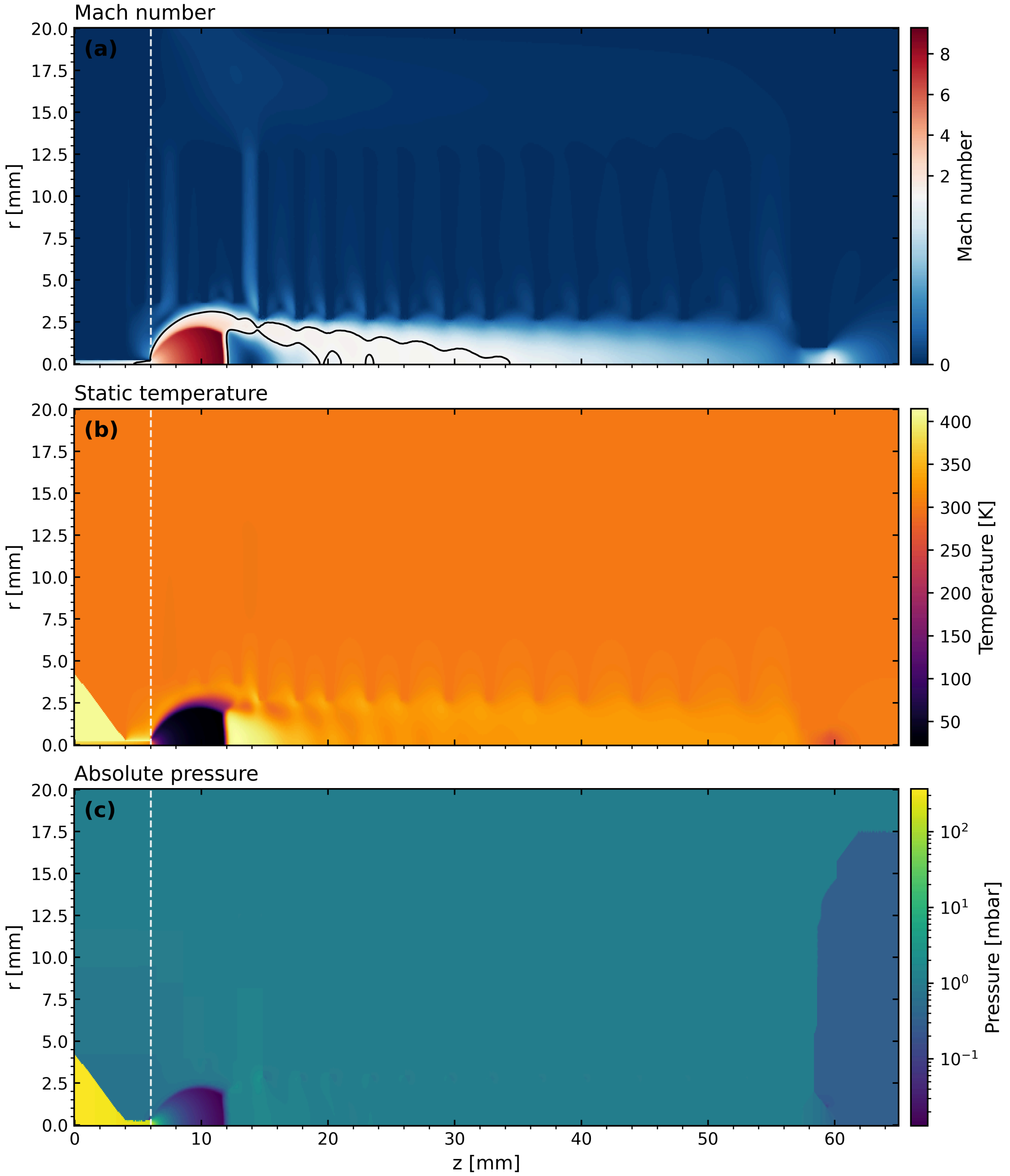

Figure S3. Imported custom-API under 1 mbar with capillary wall temperature 400K CFD diagnostics: (a) Mach number, (b) static temperature, and (c) absolute pressure. The dashed line marks the capillary-exit plane and black contours mark Mach 1.

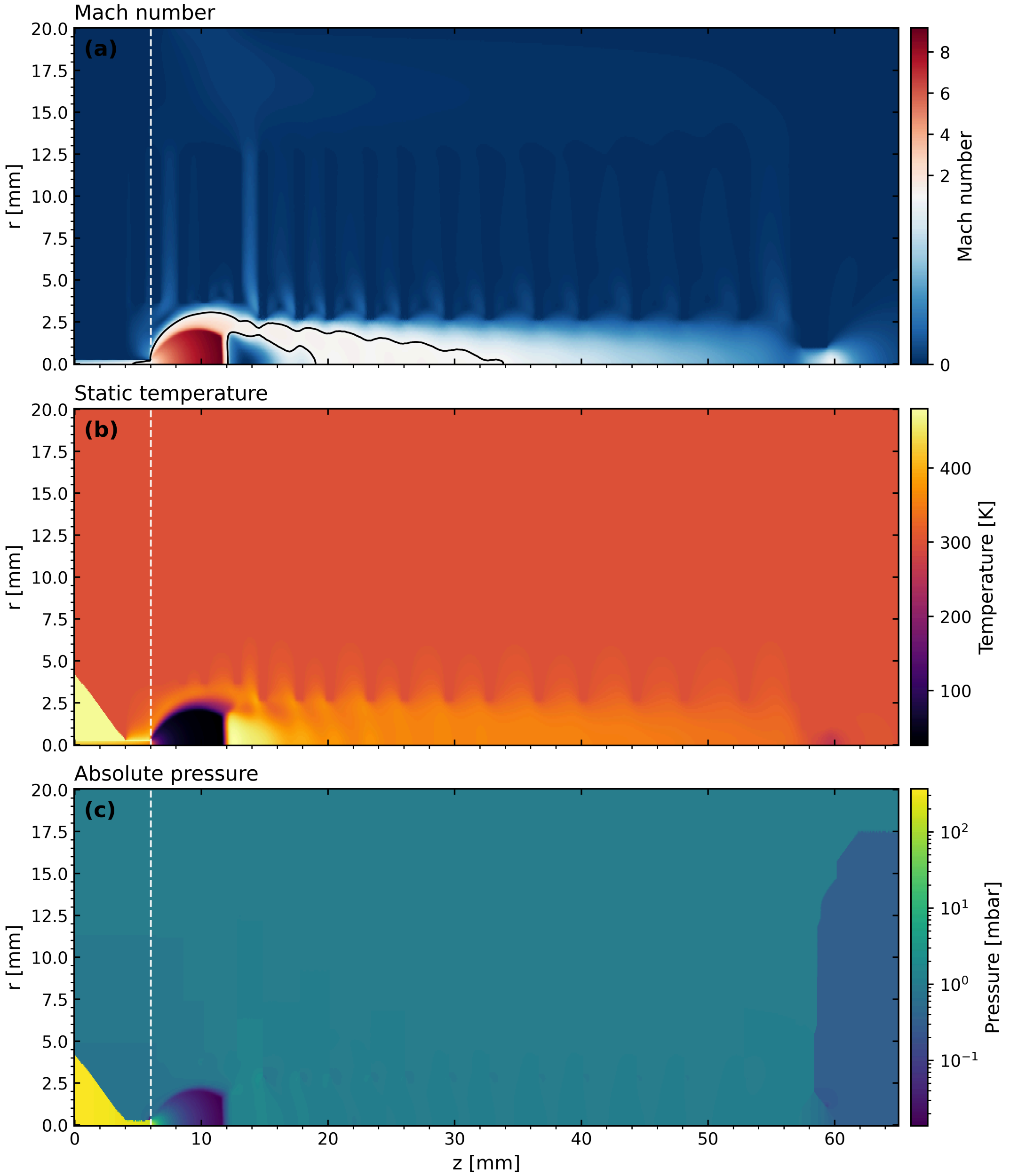

Figure S4. Imported Q-Exactive instrument source under 2 mbar with capillary wall temperature 473K CFD diagnostics: (a) Mach number, (b) static temperature, and (c) absolute pressure. The dashed line marks the capillary-exit plane and black contours mark Mach 1.

# S2. IICT backend validation

Three configurations were examined: the IonSPA 1.0.1f; IICT-lite with the same internal-energy–temperature relation $U(T)$ (taken from ref[1]); and IICT-lite with matched thermal, pseudoatom mass, and dissociation rate parameterizations[1]. The shared $U(T)$ table was reconstructed from the production peptide heat-capacity profile. The second configuration retained a fixed pseudoatom mass of 30 Da and the IICT-lite Eyring-rate implementation. In the fully matched configuration, the state-dependent pseudoatom mass and dissociation rate were supplied through a diagnostic adapter to IonSPA, while IICT-lite retained its independent collision core. Random seeds and Gaussian sampling order were matched between backends.

Each simulation followed 512 protonated leucine-enkephalin ion packs for 2 μs, starting at an internal temperature of 800 K. The ion parameters were (m/z = 556.27), charge state +1, 78 atoms, and a collision cross section of 230 $Å^2$. Uniform nitrogen gas was specified at 2 mbar and 300 K, with an axial velocity of 100 m/s. These results verify consistency of the collision implementations under matched inputs, IICT-lite with matched thermal, pseudoatom mass, and dissociation rate parameterizations was used in the simulation.

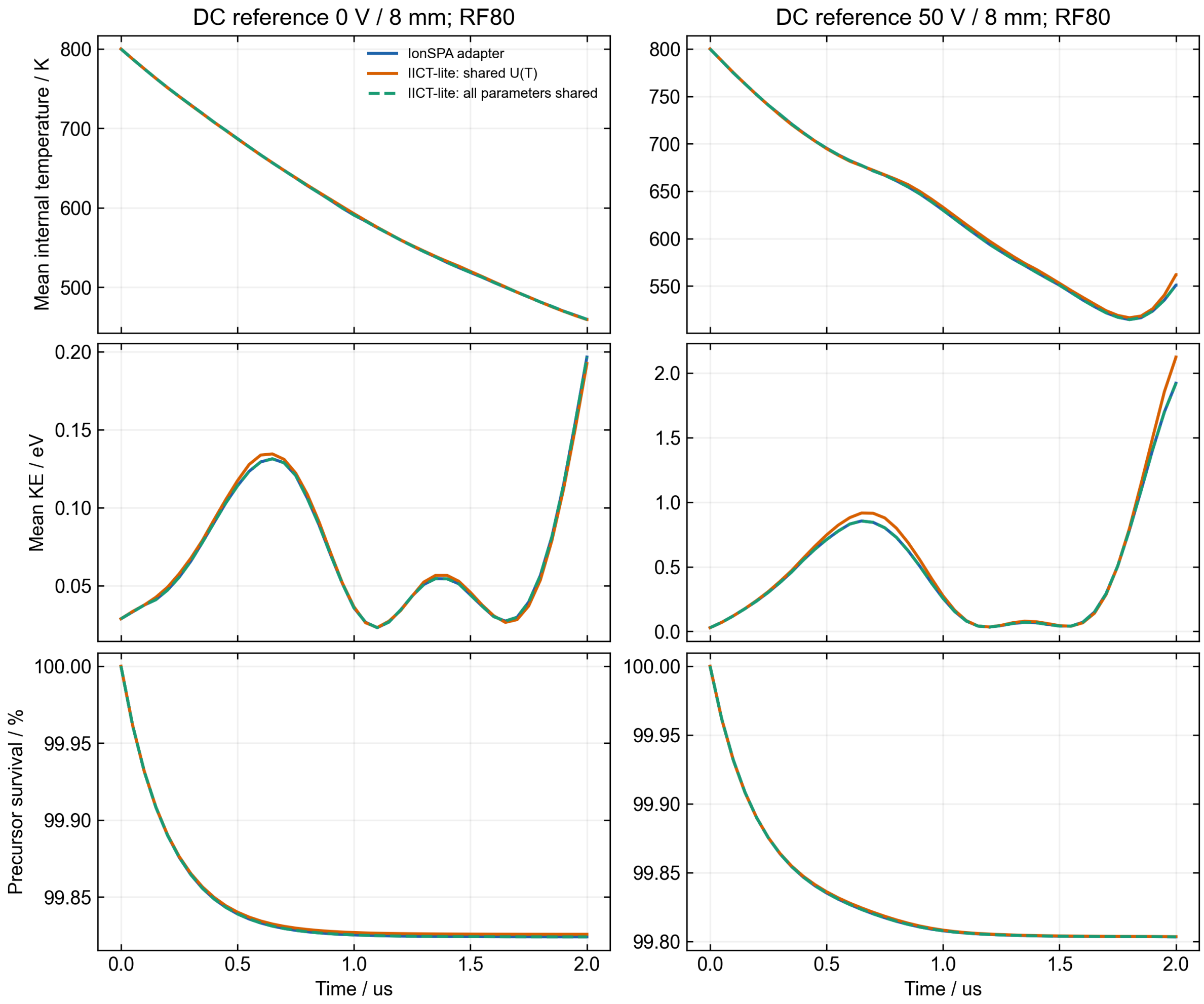

Figure S5. Matched-parameter comparison of the MuCITE IonSPA adapter and IICT-lite in prescribed RF/DC fields. Rows show mean internal temperature, mean kinetic energy, and precursor survival. Blue curves represent the IonSPA adapter, orange curves IICT-lite with shared $U(T)$ only, and green dashed curves IICT-lite with fully matched parameterizations. The green dashed and blue curves overlap.

# S3. Pure-field SIMION verification

The strict field-path benchmark uses the same 1000 initial rays, coordinate mapping, RF phase convention, electrode mask, and detector plane in SIMION and MuCITE. Gas forces, collisions, and space charge are disabled. Terminal classifications and particle IDs are compared before continuous-current or multiphysics conclusions are evaluated.

Table S1. Particle-by-particle pure-field benchmark.

| Metric | SIMION | MuCITE |
|---|---|---|
| Transmitted / outside work bench | 989 | 989 |
| Electrode hits | 11 | 11 |
| Matched hit ion IDs | 11/11 | 11/11 |
| Absolute TOF error (95% below) | Reference | 0.00210 µs |
| Absolute KE error (95% below) | Reference | 0.00892 eV |

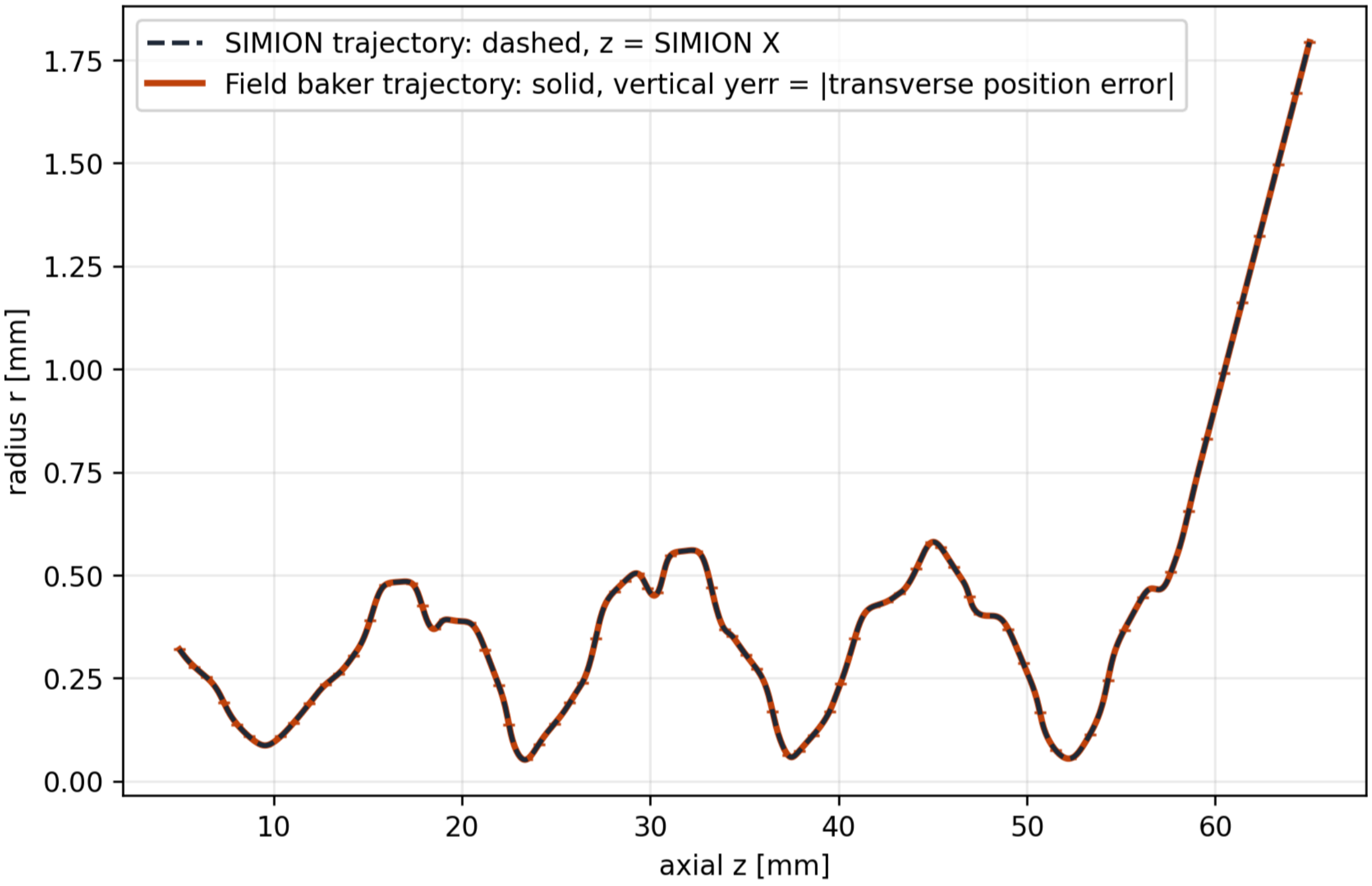

Figure S6. Representative radial trajectory overlay for the matched SIMION and baked-field MuCITE benchmark. The full verification is based on terminal classification and particle-level TOF/KE errors for 1000 matched rays, not on this single displayed trajectory alone.

# S4. Production simulation parameters

The Rhodamine B heat-capacity profile and activation parameters are empirical proxy inputs used in the production commands, not molecule-specific thermochemical measurements. Consequently, Rhodamine B dissociation fractions are interpreted as parameter-dependent activation propensity.

Table S2. Shared numerical parameters.

| Parameter | Production value |
|---|---|
| Particle backend | Taichi |
| Random seed | 7 |
| Macro-particle capacity | 120000 |
| Target represented weight | 1000 real ions per macro-ion |
| PIC update interval | $5 \times 10^{-8}$ s |
| Static field grid | $2001 \times 6501$; dr = dz = 0.01 mm |
| PIC grid | $201 \times 651$; dr = dz = 0.10 mm |
| Poisson solver | Ruge-Stüben AMG-preconditioned CG |
| Poisson tolerance / max iterations | $10^{-6}$ / 150 |
| IICT backend | IICT-lite |
| RF frequency / phase | 650 kHz / −90° |
| Detector plane / radius | 61 mm / 20 mm |
| Electrode-hit distance | 0.01 mm |
| Hybrid switch / maximum step | $\lambda\Delta t \geq 0.05$ / 2 ns |

Table S3. Case-specific physical and run parameters.

| Parameter | Custom API: Rhodamine B | Q Exactive: Leucine Enkephalin |
|---|---|---|
| Injected current | 0.5-6 nA current sweep; 2 nA RF sweep | 2 nA |
| Nominal chamber pressure | 1 mbar | 2 mbar |
| Capillary-wall temperature | 400 K | 473 K |
| Ion mass | 443.23 u | 556.27 u |
| Charge state | +1 | +1 |
| CCS in $N_2$ | $2.07 \times 10^{-18}$ $m^2$ (Ref[2]) | $2.30 \times 10^{-18}$ $m^2$ (Ref[3]) |
| Atom count | 64 | 78 |
| Heat-capacity profile | drags[1] | peptide[1] |
| $\Delta H^{\ddagger}$ | 100 $kJ \cdot mol^{-1}$ | 111.7 $kJ \cdot mol^{-1}$ (Ref[4]) |
| $\Delta S^{\ddagger}$ | −80 $J \cdot mol^{-1} \cdot K^{-1}$ | −38.1 $J \cdot mol^{-1} \cdot K^{-1}$ (Ref[4]) |
| Initial $T_{eff}$ | 300 K | 300 K |
| Collision model | Hybrid - Langevin (z = 6-7 mm) with explicit IICT outside | Hybrid - Langevin (z = 6-7 mm) with explicit IICT outside |
| Capillary voltage relative to field reference | 20 V | 0 V |
| RF amplitude | 100 $V_{peak}$ current sweep; 0-160 $V_{peak}$ RF sweep | 30,50,70,80,90,100 V |
| DC-bias series | 0 V | 0-100 V in 10 V steps |
| Simulation duration | 1 ms | 1 ms |
| Stable analysis window | 0.5-1 ms | 0.5-1ms |

## S5. Time resolved ion transfer

Ions require a certain amount of time to travel a distance. Figure S3 shows the ion exit status of the Rhodamine B ion every 0.2 ms. It is shown that in the initial 0 - 0.2 ms, ions are still flying inside the ion source, with only a small number passing through the z-axis exit. During 0.2 - 0.6 ms, ions continuously pass through the exit, and the number of ions gradually increases. After 0.6 ms, the exit current tends to stabilize.

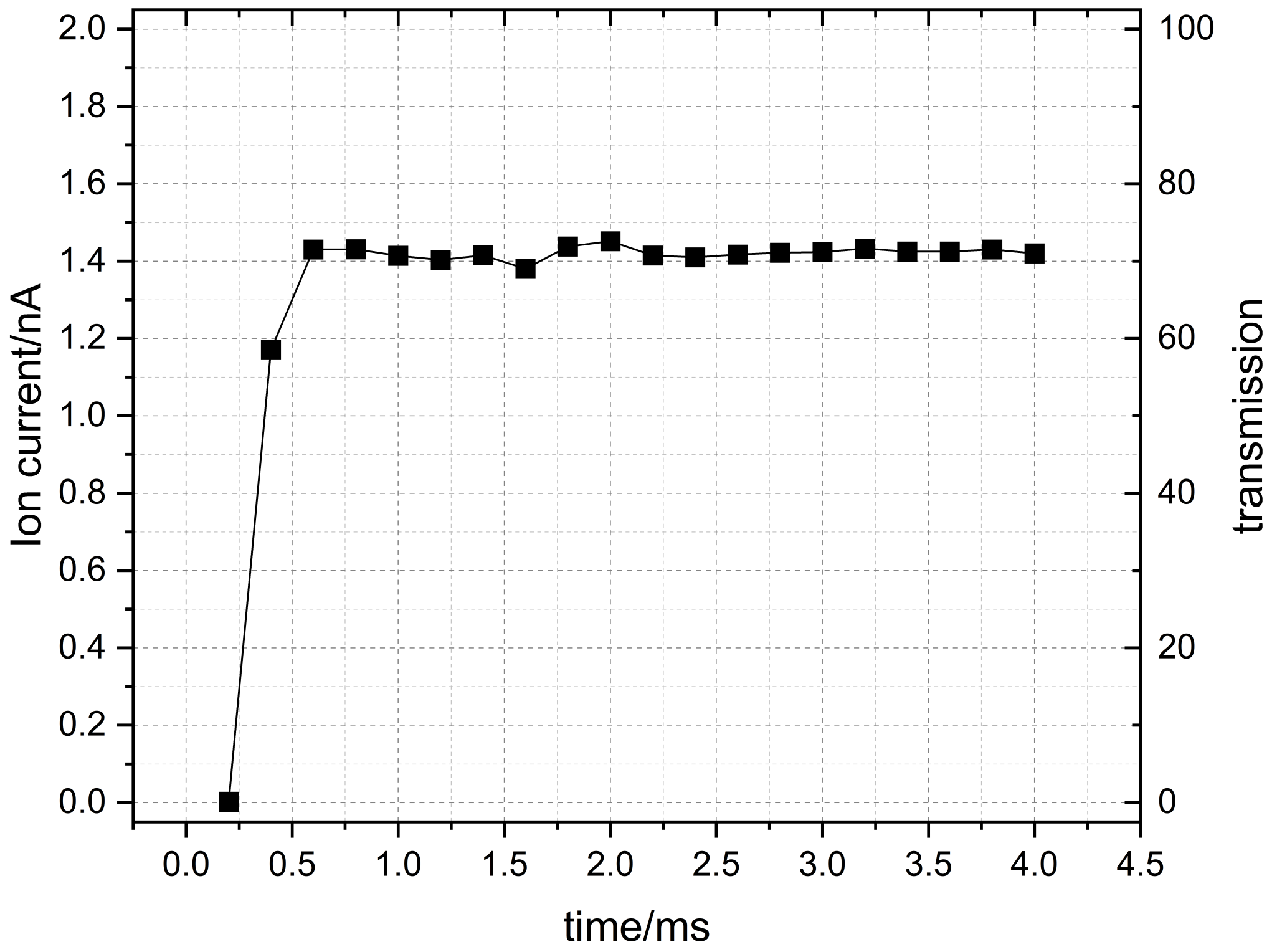

Figure S7. Ion current and transmission were calculated by weighting ions over time through the z-exit. 2 nA Rhodamine B was used with a RF amplitude of 100 $V_{peak}$.

# S6. Current-dependent exit distributions

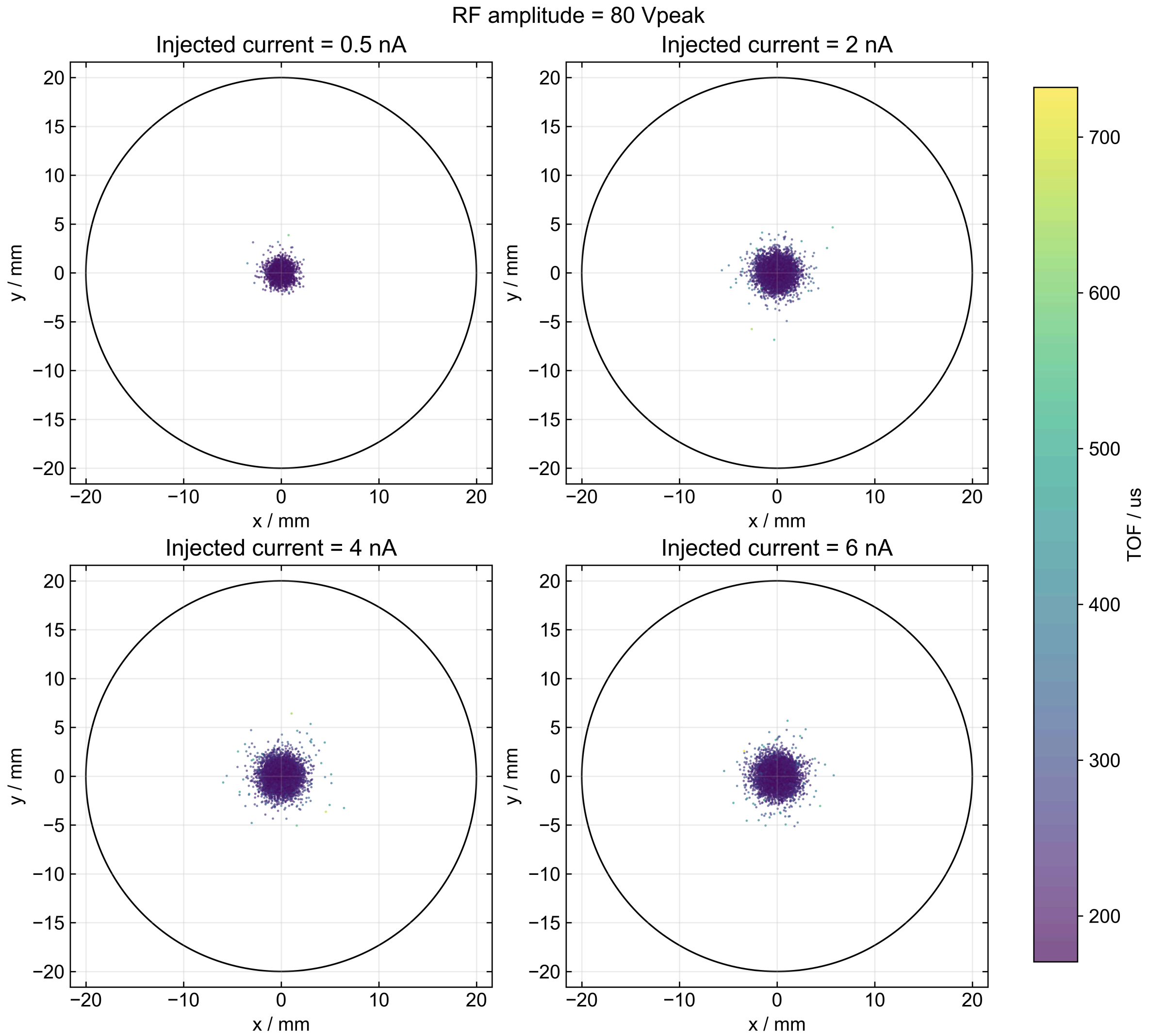

Figure S8. Z-exit x-y distributions at 0.5, 2, 4, and 6 nA for the 80 V custom-API current sweeps. Points are colored by time of flight. The 20 mm circle is the configured detector radius, not the exit-lens aperture. Increasing current broadens the transmitted population, consistent with the increase in the weighted radius enclosing 95% of the exit population reported in the main text.

## S7. Comparison of explicit and hybrid Langevin modes

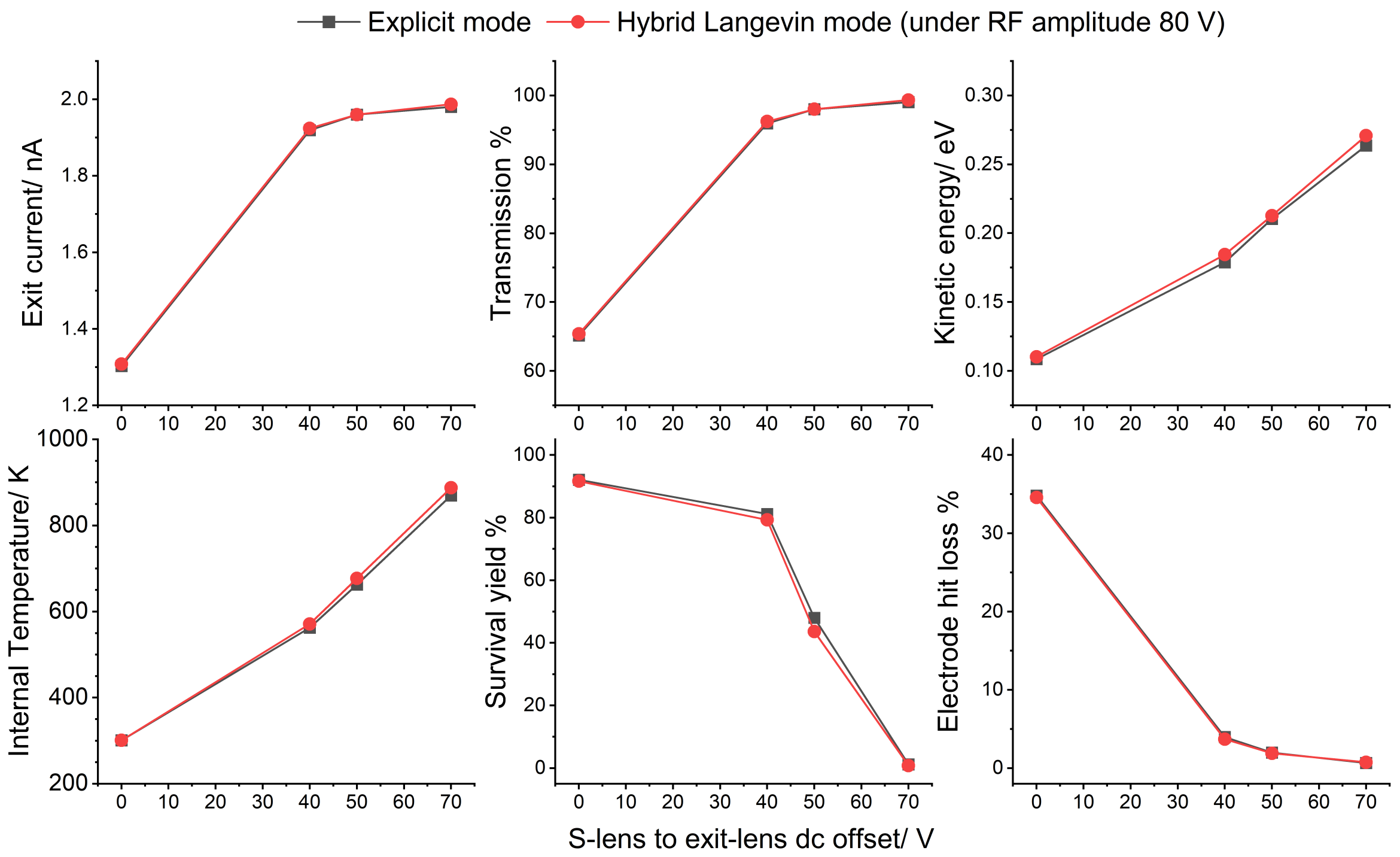

Figure S9. Comparison of explicit and hybrid Langevin modes as a function of the DC offset between the S-lens and exit lens. Exit current, transmission, kinetic energy, internal temperature, survival yield, and electrode-hit loss are shown at an RF amplitude of 80 V. Gray squares and red circles represent the explicit and hybrid Langevin modes, respectively. The two modes show close agreement across the investigated DC-offset range.

# Part II: Mathematical Methods

## S8. Symbols and units

Ion motion is treated in the nonrelativistic, quasi-electrostatic limit. Applied fields and the particle-in-cell (PIC) space-charge field are defined on axisymmetric grids, while particle positions and velocities retain all three Cartesian components. The gas flow is prescribed, and ion feedback on the flow is neglected. Improved Impulsive Collision Theory (IICT) represents momentum and internal-energy exchange through an effective pseudo-atom. Temperature-dependent first-order kinetics describe precursor loss, reported as the predicted dissociated precursor fraction or survival yield.

| Symbol | Definition | Internal unit |
|---|---|---|
| $\boldsymbol{x}, \boldsymbol{v}$ | Ion position and laboratory-frame velocity | m, m/s |
| $M, q = z_q e$ | Single-ion mass and charge | kg, C |
| $W, W_p, W_d$ | Transport, precursor, and dissociation-tagged weights | Number of represented ions |
| $r, z$ | Radial and axial coordinates | m |
| $\varphi, \boldsymbol{E}$ | Electric potential and field | V, V/m |
| $\rho_q$ | PIC charge density | $\mathrm{C/m^3}$ |
| $P_g, T_g, n_g, \rho_g$ | Gas static pressure, static temperature, number density, and mass density | Pa, K, $\mathrm{m^{-3}}$, $\mathrm{kg/m^3}$ |
| $\boldsymbol{u}_g, m_g$ | Bulk gas velocity and neutral molecular mass | m/s, kg |
| $\sigma_c$ | Effective ion-gas collision cross section | $\mathrm{m^2}$ |
| $\sigma_s, R_s$ | Transverse Gaussian source width and truncation radius | m |
| $U, T_i$ | Internal energy per ion and effective internal temperature | J/ion, K |
| $k(T_i), \mathcal{H}$ | Dissociation rate constant and cumulative hazard | $\mathrm{s^{-1}}$, dimensionless |
| $\Delta H^\ddagger, \Delta S^\ddagger$ | Activation enthalpy and entropy | J/mol, $\mathrm{J \cdot (mol \cdot K)^{-1}}$ |
| $k_B, h, N_A, R$ | Boltzmann, Planck, Avogadro, and gas constants | SI; $R = N_A k_B$ |

The source width and collision cross section are denoted by $\sigma_s$ and $\sigma_c$, respectively. The cumulative dissociation hazard is $\mathcal{H}$. The effective internal temperature $T_i$, denoted by $T_{eff}$ in the figures, represents molecular internal energy; $T_g$ is the gas static temperature.

The unit conversions are $1\ \mathrm{mm} = 10^{-3}\ \mathrm{m}$, $1\ \mathrm{mbar} = 100\ \mathrm{Pa}$, and $1\ \mathrm{eV} = 1.602176634 \times 10^{-19}\ \mathrm{J}$.

# S9. Weighted ion packs and the continuous-current source

## S9.1. From physical ions to computational particles

A computational ion pack represents $W_j$ physical ions with the same representative state. Following the weighted-particle representation used in particle simulations[5], the ion distribution is approximated by the empirical measure

$$f_N(\boldsymbol{x}, \boldsymbol{v}, U, t) \simeq \sum_j W_j \delta(\boldsymbol{x} - \boldsymbol{x}_j)\delta(\boldsymbol{v} - \boldsymbol{v}_j)\delta(U - U_j). \tag{S1}$$

Integration over phase space gives the represented ion number, $N = \sum_j W_j$, and total charge, $Q = \sum_j q_j W_j$. Scaling both pack mass and charge by $W$ leaves the electric acceleration equal to $\frac{q\boldsymbol{E}}{M}$. Particle motion therefore uses the single-ion charge-to-mass ratio, whereas PIC deposition uses $qW$.

Each pack is a weighted sample of the ion phase-space distribution. Its collision probability is calculated from the local state of a single ion. One random collision update changes the representative pack state, and an ensemble of packs samples the ion population.

## S9.2. Current and injection rate

Current is the charge crossing a surface per unit time. For a source with a single charge state and a positive current magnitude,

$$I_{\mathrm{in}} = |q| \frac{dN}{dt}, \quad \dot{N} = \frac{I_{\mathrm{in}}}{|q|}. \tag{S2}$$

For a target pack weight $W_*$, the expected injection rate is $\dot{N}_{\mathrm{pack}} = I_{\mathrm{in}}/(|q|W_*)$. A singly charged 2 nA source with $W_* = 1000$, for example, injects approximately $1.2483 \times 10^7$ packs/s. The current sets the injected charge per unit time, while the target pack weight sets the number of computational packs used to represent it.

An accumulator stores the represented ions that have not yet formed a complete pack. If $A_n$ is the balance before a step,

$$A^* = A_n + \frac{I_{\mathrm{in}}\Delta t}{|q|}, \quad n_{\mathrm{emit}} = \left\lfloor \frac{A^*}{W_*} \right\rfloor, \quad A_{n+1} = A^* - n_{\mathrm{emit}} W_*. \tag{S3}$$

Because $0 \le A_{n+1} < W_*$, the cumulative injection error is less than one pack weight when slots are available. Complete packs that cannot be accommodated are recorded as blocked source injection, reducing the realized current.

A positive `ion_pack_weight` specifies the target weight directly. Otherwise, the target is obtained by dividing the nominal macro-step injection by `ion_pack_per_injection`. The `ion_count` parameter specifies the capacity of the reusable slot pool.

## S9.3. Truncated transverse Gaussian distribution

Independent coordinates $x, y \sim \mathcal{N}(0, \sigma_s^2)$ are drawn, and points with $r = \sqrt{x^2 + y^2} > R_s$ are rejected. The normalized two-dimensional density is

$$p(x,y) = \frac{\exp\left[-\frac{x^2+y^2}{2\sigma_s^2}\right]}{2\pi\sigma_s^2\left[1 - \exp\left(-\frac{R_s^2}{2\sigma_s^2}\right)\right]}, \quad x^2 + y^2 \leq R_s^2. \tag{S4}$$

Using the area element $dxdy = rdrd\theta$ and integrating over the azimuthal angle gives the radial density and cumulative distribution,

$$p_r(r) = \frac{r\exp\left[-\frac{r^2}{2\sigma_s^2}\right]}{\sigma_s^2\left[1 - \exp\left(-\frac{R_s^2}{2\sigma_s^2}\right)\right]}, \quad F_r(r) = \frac{1 - \exp\left[-\frac{r^2}{2\sigma_s^2}\right]}{1 - \exp\left[-\frac{R_s^2}{2\sigma_s^2}\right]}. \tag{S5}$$

Because annular area is proportional to $r$, equal-width radial bins contain fewer samples near the axis, even though the density per unit area is highest there. The Gaussian source uses rejection sampling. The uniform-disk option instead follows $F_r = r^2/R_s^2$, giving $r = R_s\sqrt{\xi}$ for a uniform random variable $\xi$.

In the study cases, $\sigma_s = 0.20$ mm is the standard deviation of each transverse Cartesian component before truncation, and $R_s = 0.25$ mm is the injection boundary corresponding to the capillary inner radius.

## S9.4. Initial velocity and the capillary buffer

Initial velocities are sampled from the original CFD data. An axial coordinate is drawn uniformly within a specified interval; the local bulk gas velocity is interpolated at the birth radius, and its radial component is projected onto the $x$ and $y$ directions. Nearest-neighbor values are used outside the scattered-data convex hull. The birth velocity is set to this bulk velocity; neutral thermal motion is sampled in the collision calculation. An optional Mach-number source sets $v_z = M_a\sqrt{(\gamma R T_g)/M_g^{\mathrm{mol}}}$.

The source-acceleration parameter `capillary_voltage_v` modifies the birth kinetic energy. For the axial-velocity branch,

$$K_{z,\mathrm{new}} = \max\left(\frac{1}{2}Mv_z^2 + |q|\ V_{\mathrm{cap}}, 0\right), \quad |v_{z,\mathrm{new}}| = \sqrt{\frac{2K_{z,\mathrm{new}}}{M}}. \tag{S6}$$

A positive correction directs the axial velocity downstream, while a negative correction preserves its original sign. The buffer can be prefilled with the equivalent weight $N_{\mathrm{buffer}} \simeq \left(\dot{N} L_{\mathrm{buffer}}\right)/v_z$, obtained from particle flux multiplied by residence time. Buffered particles move according to $z \leftarrow z + v_z \Delta t$ and enter the transport population when they cross the capillary exit. Before activation, they are excluded from PIC charge deposition, electric-field RK4 propagation, and dissociation accumulation.

# S10. Precomputed electric fields and three-dimensional coordinates

## S10.1. Potential superposition and RF convention

With prescribed electrode potentials and space charge treated separately, the applied electrode potentials satisfy Laplace's equation. Its linearity allows independent excitations to be superposed:

$$\nabla^2 \varphi_{\mathrm{DC}} = 0, \quad \nabla^2 \varphi_{\mathrm{RF,ref}} = 0, \tag{S7}$$

$$\varphi_{\mathrm{ext}}(r, z, t) = \varphi_{\mathrm{DC}}(r, z) + \frac{V_{\mathrm{peak}}}{V_{\mathrm{ref}}} \varphi_{\mathrm{RF,ref}}(r, z) \cos(\omega t + \varphi), \quad \omega = 2\pi f. \tag{S8}$$

Taking the negative gradient gives the corresponding field superposition. Adjacent electrodes in the S-lens RF basis are assigned $+1$ V and $-1$ V, with $V_{\mathrm{ref}} = 1\mathrm{V}_{\mathrm{peak}}$. RF amplitudes in this document denote the peak voltage of one electrode relative to the reference potential, $V_{\mathrm{peak}}$. Both the single-electrode peak-to-peak voltage and the peak voltage difference between opposite-phase neighbors are $2V_{\mathrm{peak}}$.

Space charge is added as a perturbation field:

$$\boldsymbol{E}_{\mathrm{total}}(\boldsymbol{x}, t) = \boldsymbol{E}_{\mathrm{DC}} + s_{\mathrm{RF}}(t) \boldsymbol{E}_{\mathrm{RF,ref}} + s_{\mathrm{SC}} \boldsymbol{E}_{\mathrm{SC}}, \quad s_{\mathrm{RF}}(t) = \frac{V_{\mathrm{peak}}}{V_{\mathrm{ref}}} \cos(\omega t + \varphi). \tag{S9}$$

The applied DC and RF fields superpose linearly, although the resulting trajectories, collisions, and dissociation responses may be nonlinear.

## S10.2. From axisymmetric fields to three-dimensional forces

An axisymmetric scalar potential is independent of the azimuthal angle. Applying the chain rule, $\frac{\partial r}{\partial x} = \frac{x}{r}$, gives

$$r = \sqrt{x^2 + y^2}, \quad E_x = \frac{E_r x}{r}, \quad E_y = \frac{E_r y}{r}, \quad E_z = E_z(r, z). \tag{S10}$$

Symmetry sets the transverse field components to zero on the axis. The axisymmetric field therefore supplies three-dimensional forces while particle propagation retains the full transverse velocity direction and azimuthal motion.

# S11. Gas fields and thermal motion

## S11.1. Ideal-gas closure

Dividing the ideal-gas equation $PV = Nk_BT$ by volume gives[6]

$$n_g = \frac{P_g}{k_B T_g}, \quad m_g = \frac{M_g^{\mathrm{mol}}}{N_A}, \quad \rho_g = n_g m_g. \tag{S11}$$

Collision calculations use absolute pressure and static temperature.

Scattered Fluent data are linearly interpolated onto the common field grid. Configured background values fill points outside the convex hull or retained domain. Runtime gas sampling then uses bilinear interpolation. Shocks, boundary layers, and expansion flow enter the ion model through these precomputed CFD fields.

## S11.2. Maxwell distribution and Mach number

The velocity of a neutral molecule in local equilibrium is the sum of bulk flow and a thermal fluctuation:

$$\boldsymbol{v}_g = \boldsymbol{u}_g + \boldsymbol{g}, \quad g_x, g_y, g_z \sim \mathcal{N}\left(0, \frac{k_B T_g}{m_g}\right). \tag{S12}$$

Combining and normalizing the Boltzmann factors $\exp\left[-\frac{m_g g_i^2}{2k_B T_g}\right]$ for the three Cartesian components yields the Maxwell velocity density[6],

$$f_g(\boldsymbol{g}) = \left(\frac{m_g}{2\pi k_B T_g}\right)^{\frac{3}{2}} \exp\left[-\frac{m_g\,|\boldsymbol{g}|^2}{2k_B T_g}\right]. \tag{S13}$$

The sound speed is defined by the isentropic derivative $c_s^2 = \left(\frac{\partial P}{\partial \rho}\right)_s$. For an ideal gas with constant $P\rho^{-\gamma}$,

$$c_s = \sqrt{\frac{\gamma R T_g}{M_g^{\mathrm{mol}}}}, \quad M_a = \frac{|\boldsymbol{u}_g|}{c_s}. \tag{S14}$$

The Mach number characterizes the bulk gas velocity. A high-Mach-number expansion can coexist with a low static temperature.

# S12. Axisymmetric PIC charge deposition

## S12.1. Cloud-in-cell weights

PIC alternates between particle and grid-field updates. Cloud-in-cell (CIC) assigns charge to the grid and interpolates fields back to particles[5]. The one-dimensional linear shape function is $S(\xi) =$ $\max(1-|\xi|,0)$. Let $\xi$ and $\eta$ denote the normalized radial and axial coordinates within a cell. The product of the two shape functions gives four bilinear weights:

$$w_{00}=(1-\xi)(1-\eta),\quad w_{10}=\xi(1-\eta),\quad w_{01}=(1-\xi)\eta,\quad w_{11}=\xi\eta. \tag{S15}$$

Each three-dimensional pack is mapped to $(r,z)$, and its charge is distributed to the surrounding nodes as $qWw$. In the axisymmetric representation, each pack's charge is distributed uniformly around an azimuthal ring.

### S12.2. Annular control volumes

A node at $r_i=i\Delta r$ represents the annulus bounded by $r_-=\max\bigl(0,r_i-\frac{\Delta r}{2}\bigr)$ and $r_+=\min\bigl(r_{\max},r_i+\frac{\Delta r}{2}\bigr)$. Integrating the cylindrical volume element gives

$$V_i=\int_{z_k-\frac{\Delta z}{2}}^{z_k+\frac{\Delta z}{2}}\int_{r_-}^{r_+}\int_0^{2\pi} r\,\mathrm{d}\theta\,\mathrm{d}r\,\mathrm{d}z=\pi(r_+^2-r_-^2)\Delta z. \tag{S16}$$

The axis node has volume $V_0=(\pi\Delta r^2\Delta z)/4$. Charge deposition uses the full axial width $\Delta z$ at each node; the axial endpoints are treated by the potential boundary conditions.

$$\rho_{q,i,k}=\frac{1}{V_i}\sum_p q_p W_p w_{i,k}\bigl(r_p,z_p\bigr). \tag{S17}$$

The four weights of each pack sum to unity, so summing $\rho_q V$ over the deposition nodes recovers the total deposited charge. Bilinear weights combined with annular volumes conserve charge exactly. Close to the axis, the geometric weighting can introduce small deviations when representing a uniform spatial density.

### S12.3. Macro-step update

At each macro step, the charge density is cleared, active transport packs deposit their charge, and the space-charge field is solved. This field remains fixed until the next macro step, while the applied RF field continues to vary during micro steps. Newly injected charge enters the next PIC update, introducing a finite field-update lag.

## S13. Poisson equation and finite-volume discretization

### S13.1. From Gauss's law to the cylindrical equation

Combining $\nabla\cdot\boldsymbol{E}=\rho_q/\varepsilon_0$ with $\boldsymbol{E}=-\nabla\varphi$ gives $\nabla^2\varphi=-\rho_q/\varepsilon_0$. With no azimuthal dependence,

$$\frac{1}{r}\frac{\partial}{\partial r}\left(r\frac{\partial\varphi_{\mathrm{SC}}}{\partial r}\right)+\frac{\partial^2\varphi_{\mathrm{SC}}}{\partial z^2}=-\frac{\rho_q}{\varepsilon_0}. \tag{S18}$$

The axis satisfies $\partial_r \varphi_{\mathrm{SC}} = 0$. The axial ends, outer radial boundary, and metal-mask nodes satisfy $\varphi_{\mathrm{SC}} = 0$.

### S13.2. Integrated face fluxes

Integrating over a control volume and applying the divergence theorem converts the Laplacian into normal-gradient fluxes through the faces. A two-point difference approximates each face gradient, following the finite-volume construction for elliptic equations[7]:

$$\sum_N G_{PN}(\varphi_N - \varphi_P) = -\frac{\rho_{q,P} V_P}{\varepsilon_0}. \tag{S19}$$

The radial face areas are $2\pi r_\pm \Delta z$, and the axial face area is $A_r = \pi(r_+^2 - r_-^2)$. The four flux coefficients are therefore

$$G_E = \frac{2\pi r_+ \Delta z}{\Delta r}, \quad G_W = \frac{2\pi r_- \Delta z}{\Delta r}, \quad G_N = G_S = \frac{A_r}{\Delta z}. \tag{S20}$$

Setting $A_{PP} = -\sum_N G_{PN}$, $A_{PN} = G_{PN}$, and $b_P = -(\rho_{q,P} V_P)/\varepsilon_0$ gives the sparse system $A\varphi = b$. Zero-potential boundary nodes are excluded from the unknowns, while their face coefficients remain in the diagonal entries.

On the axis, $G_W = 0$. Dividing the east-face coefficient by the axis-node volume gives $G_E/V_0 = 4/(\Delta r^2)$, and hence

$$\frac{4(\varphi_{1,k} - \varphi_{0,k})}{\Delta r^2} + \frac{\varphi_{0,k+1} - 2\varphi_{0,k} + \varphi_{0,k-1}}{\Delta z^2} = -\frac{\rho_{q,0,k}}{\varepsilon_0}. \tag{S21}$$

The same expression follows from an even Taylor expansion of a smooth axisymmetric potential, avoiding direct division by $r = 0$.

### S13.3. Symmetry and positive definiteness

In the integrated formulation, neighboring nodes share the same face coefficient. Defining $B = -A$ and $d = -b$, a nonzero vector $\boldsymbol{x}$ satisfying the homogeneous boundaries gives

$$\boldsymbol{x}^\top B \boldsymbol{x} = \sum_{\{P,N\}} G_{PN}(x_P - x_N)^2 + \sum_{P,D} G_{PD} x_P^2 > 0. \tag{S22}$$

Each undirected interior face is counted once in the first sum; the second sum represents Dirichlet faces. The free-node operator is therefore symmetric positive definite when each connected component reaches a Dirichlet boundary, allowing conjugate-gradient solution[8].

### S13.4. Electric-field recovery

The potential is differentiated with centered differences at interior nodes and one-sided differences at boundaries. The radial field is set to zero on the axis:

$$E_{r,i,k} = -\frac{\varphi_{i+1,k} - \varphi_{i-1,k}}{2\Delta r}, \quad E_{z,i,k} = -\frac{\varphi_{i,k+1} - \varphi_{i,k-1}}{2\Delta z}. \tag{S23}$$

Bilinear interpolation transfers these components to particle positions, followed by the three-dimensional projection in Section 3.

# S14. Poisson solvers and convergence criteria

## S14.1. Preconditioned conjugate gradients

Solving $B\varphi = d$ is equivalent to minimizing $J(\varphi) = \frac{1}{2}\varphi^\top B\varphi - d^\top\varphi$, whose gradient is $B\varphi - d = -r$. Conjugate search directions avoid repeatedly correcting the same error component[8]. Given an approximate inverse preconditioner $\mathcal{M}^{-1}$, the iterations are

$$r_0 = d - B\varphi_0, \quad z_0 = \mathcal{M}^{-1}r_0, \quad p_0 = z_0, \tag{S24}$$

$$\alpha_k = \frac{r_k^\top z_k}{p_k^\top B p_k}, \quad \varphi_{k+1} = \varphi_k + \alpha_k p_k, \quad r_{k+1} = r_k - \alpha_k B p_k, \tag{S25}$$

$$z_{k+1} = \mathcal{M}^{-1}r_{k+1}, \quad \beta_k = \frac{r_{k+1}^\top z_{k+1}}{r_k^\top z_k}, \quad p_{k+1} = z_{k+1} + \beta_k p_k. \tag{S26}$$

The step length $\alpha$ minimizes $J$ along the search direction; $\beta$ maintains $B$-conjugacy. Standard PCG uses a symmetric positive-definite preconditioner compatible with $B$.

## S14.2. Algebraic multigrid

For an approximate solution $\tilde{\varphi}$, the residual $r = d - B\tilde{\varphi}$ defines the error equation $Be = r$. Local smoothing reduces oscillatory error efficiently, while algebraically smooth error is treated on coarse levels. AMG selects coarse variables and interpolation $P$ from matrix connectivity and forms the Galerkin system[9]:

$$B_c = RBP, \quad r_c = Rr, \quad B_c e_c = r_c, \quad \tilde{\varphi} \leftarrow \tilde{\varphi} + Pe_c. \tag{S27}$$

Here, $R$ is usually $P^\top$. A V-cycle recursively applies presmoothing, residual restriction, coarse solution, prolongation, and postsmoothing. The Ruge-Stueben method constructs the algebraic hierarchy from strong matrix connections. Other options include smoothed aggregation, standalone AMG iteration, and AMG-preconditioned BiCGSTAB. The implementation uses PyAMG[10].

For a fixed PIC grid and metal mask, the matrix and AMG hierarchy are built once; each macro step updates only the right-hand side. Warm starts use the preceding potential as the initial guess without changing the convergence criterion. A sparse direct solver can instead factorize $A = LU$ and solve two triangular systems, reusing the factorization while the matrix remains fixed.

### S14.3. Successive over-relaxation

The node equation gives a candidate potential $\varphi_P^*$, followed by relaxation[11]:

$$\varphi_P^* = \frac{\sum_N G_{PN}\varphi_N + \frac{\rho_{q,P} V_P}{\varepsilon_0}}{\sum_N G_{PN}}, \quad \varphi_P^{k+1} = (1-\omega_{\text{SOR}})\varphi_P^k + \omega_{\text{SOR}}\varphi_P^*. \tag{S28}$$

Nodes are divided into red and black sets according to $(i+k)$ mod 2 and updated alternately, with $0 < \omega_{\text{SOR}} < 2$.

### S14.4. Convergence criterion

The residual of the original linear system is recomputed, using

$$\varepsilon_{\text{rel}} = \frac{\|b - A\varphi\|_2}{\|b\|_2} \leq \varepsilon_{\text{tol}}. \tag{S29}$$

When the right-hand-side norm is close to zero, the denominator is replaced by the larger of the initial residual norm and a small positive floor. AMG/Krylov solutions must also be finite and return a successful solver status. In SOR, the maximum node update triggers residual evaluation; the relative residual determines final convergence.

## S15. Equations of motion and RF heating

### S15.1. Deterministic propagation

Neglecting gravity, Newton's equations are

$$\frac{d\boldsymbol{x}}{dt} = \boldsymbol{v}, \quad \frac{d\boldsymbol{v}}{dt} = \frac{q}{M}\boldsymbol{E}_{\text{total}}(\boldsymbol{x}, t). \tag{S30}$$

The stochastic collision operator handles gas-induced momentum exchange. Differentiating the translational kinetic energy, $K = M|\boldsymbol{v}\frac{|^2}{2}$, gives the electric power transferred to the ion:

$$\frac{dK}{dt} = M\boldsymbol{v} \cdot \frac{d\boldsymbol{v}}{dt} = q\boldsymbol{E} \cdot \boldsymbol{v}. \tag{S31}$$

In a static conservative field without collisions, integration gives $K_f - K_i = q\big(\varphi_i - \varphi_f\big)$. During coupled transport, both DC and RF fields do work, and collisions redistribute energy to the gas and molecular internal degrees of freedom. Exit kinetic energy is therefore determined by the field work and collision history along the trajectory.

### S15.2. RF micromotion and pseudopotential

For a locally constant RF amplitude, solving $M\ddot{\xi} = qE_0\cos(\omega t)$ gives the micromotion $\xi = -\frac{qE_0\cos(\omega t)}{M\omega^2}$ and oscillatory velocity $\dot{\xi} = \frac{qE_0\sin(\omega t)}{M\omega}$. Averaging over one period gives[12]

$$\langle K_{\mathrm{micro}} \rangle = \frac{q^2 E_0^2}{4M\omega^2}. \tag{S32}$$

If the amplitude varies slowly in space, expanding $E_0(X+\xi)$ to first order and averaging $q\xi \nabla E_0 \cos \omega t$ gives the pseudopotential energy and effective force:

$$\Psi_{\mathrm{RF}} = \frac{q^2 \, |\boldsymbol{E}_0|^2}{4M\omega^2}, \quad \boldsymbol{F}_{\mathrm{eff}} = -\nabla \Psi_{\mathrm{RF}}. \tag{S33}$$

These expressions apply in the adiabatic, slowly varying-amplitude limit. MuCITE resolves the time-dependent RF field directly and uses the collision model to calculate energy exchange involving micromotion, translation, and internal excitation.

# S16. RK4 and macro/micro time splitting

## S16.1. Fourth-order Runge-Kutta integration

Writing the state as $y = (\boldsymbol{x}, \boldsymbol{v})$ with $\dot{y} = F(t, y)$, the classical RK4 scheme samples four slopes to match the Taylor expansion through fourth order[13]:

$$k_1 = F(t_n, y_n), \quad k_2 = F\left(t_n + \frac{\Delta t}{2}, y_n + \frac{\Delta t k_1}{2}\right), \tag{S34}$$

$$k_3 = F\left(t_n + \frac{\Delta t}{2}, y_n + \frac{\Delta t k_2}{2}\right), \quad k_4 = F(t_n + \Delta t, y_n + \Delta t k_3), \tag{S35}$$

$$y_{n+1} = y_n + \frac{\Delta t}{6}(k_1 + 2k_2 + 2k_3 + k_4). \tag{S36}$$

The weights satisfy the Runge-Kutta order conditions: $b^\top \mathbf{1} = 1$, $b^\top c = \frac{1}{2}$, $b^\top c^2 = \frac{1}{3}$, $b^\top Ac = \frac{1}{6}$, $b^\top c^3 = \frac{1}{4}$, $b^\top CAc = \frac{1}{8}$, $b^\top Ac^2 = \frac{1}{12}$, and $b^\top A^2 c = \frac{1}{24}$. Here, $c = \left(0, \frac{1}{2}, \frac{1}{2}, 1\right)$, $b = \frac{1,2,2,1}{6}$, $A$ is the stage-coefficient matrix, and $C = \mathrm{diag}(c)$. For smooth deterministic $F$, the local truncation error is $O(\Delta t^5)$ and the global error is $O(\Delta t^4)$.

Each stage evaluates the RF field at its own position and time. The PIC field remains fixed within the macro step.

## S16.2. Micro-step constraints

Physical and geometric constraints determine the micro-step length, which is then used for an RK4 advance. In explicit collision mode, the principal candidates are

$$\Delta t = \min_j \left\{ \Delta t_{\max}, \Delta t_{\mathrm{remain}}, \frac{c_{\mathrm{RF}}}{f}, \frac{c_{\mathrm{coll}}}{\nu_j}, c_a \sqrt{\frac{2\ell_{\mathrm{tol}}}{\frac{|q_j| \, |\boldsymbol{E}_j|}{M_j}}}, \frac{c_{\mathrm{CFL}} h_{\min}}{|\boldsymbol{v}_j|} \right\}. \tag{S37}$$

The RF constraint resolves each period with at least $\frac{1}{c_{\mathrm{RF}}}$ steps; $c_{\mathrm{RF}} = 0.05$ gives at least 20 steps. The collision constraint enforces $\nu\Delta t \leq c_{\mathrm{coll}}$. For $c_{\mathrm{coll}} = 0.1$, the probability of at least one event is bounded by $1 - \exp(-0.1) = 9.52\%$.

The acceleration constraint follows by limiting the constant-acceleration displacement $\frac{a\Delta t^2}{2}$ to a prescribed length scale and applying the safety factor $c_a$; the resulting displacement limit is $c_a^2 \ell_{\mathrm{tol}}$.

The CFL-type constraint limits translation to a fraction of the smallest active field or mask spacing, usually with $c_{\mathrm{CFL}} = 0.5$.

The global step is the minimum candidate across all active packs, so a small population in a high-pressure region can set the propagation cost. The remaining-time constraint also prevents crossing a macro-step boundary, simulation end, or operating-stage change.

The hybrid path selects a local branch from the spatial window, the $\nu\Delta t$ threshold, and the Langevin outer-step limit, then coordinates the global step. Branch eligibility is reevaluated if the coordinated step changes these conditions.

### S16.3. Splitting error

Let $L$ denote electric-field transport and $C$ the collision generator. Applying collisions before propagation gives the first-order product $\exp(\Delta t L)\exp(\Delta t C)$. The Baker-Campbell-Hausdorff expansion is[14]

$$\exp(\Delta t L)\exp(\Delta t C) = \exp\left[\Delta t(L + C) + \frac{\Delta t^2}{2}[L, C] + O(\Delta t^3)\right]. \tag{S38}$$

The combined splitting is generally first order, although the smooth electric-field subproblem uses fourth-order integration. PIC field freezing and terminal-event localization introduce additional discretization errors. Convergence is assessed from exit flux, ion energies, and precursor survival as the relevant time steps are refined.

## S17. Collision frequency and event probability

### S17.1. Swept-volume argument

For a fixed relative speed, a collision cross section $\sigma_c$ sweeps out the volume $\sigma_c g dt$ during $dt$. Multiplying by the neutral number density and averaging over neutral thermal velocities gives[15]

$$\nu = n_g \sigma_c \bar{g}, \quad \bar{g} = \int_{\mathbb{R}^3} |\boldsymbol{w} - \boldsymbol{g}| \; f_g(\boldsymbol{g}) \mathrm{d}^3 g, \quad \boldsymbol{w} = \boldsymbol{v} - \boldsymbol{u}_g. \tag{S39}$$

The effective total collision cross section $\sigma_c$ is taken to be independent of speed.

## S17.2. Thermally averaged relative speed

Define $a = \sqrt{(2k_B T_g)/m_g}$, $s = |\boldsymbol{w}|/a$, and the relative-velocity vector $\boldsymbol{c} = \boldsymbol{w} - \boldsymbol{g}$. Choosing $\boldsymbol{w}$ as the polar axis, angular integration of the shifted Gaussian gives

$$\overline{g} = \frac{4}{\sqrt{\pi}a^3} \int_0^\infty c^3 e^{-\frac{c^2+w^2}{a^2}} \frac{\sinh\left(\frac{2cw}{a^2}\right)}{\frac{2cw}{a^2}} \, \mathrm{d}c. \tag{S40}$$

Using $2\sinh x = e^x - e^{-x}$ separates the integral into Gaussian terms in $c \pm w$. Evaluation with the error function yields

$$\overline{g} = a\left[\left(s + \frac{1}{2s}\right) \operatorname{erf}(s) + \frac{e^{-s^2}}{\sqrt{\pi}}\right]. \tag{S41}$$

Expanding the error function and exponential as $s \to 0$ removes the apparent $\frac{1}{s}$ singularity:

$$\overline{g} = \frac{2a}{\sqrt{\pi}}\left(1 + \frac{s^2}{3} - \frac{s^4}{30} + \frac{s^6}{210} - \frac{s^8}{1512} + \frac{s^{10}}{11880} + \cdots\right). \tag{S42}$$

At zero drift, $\overline{g} = \sqrt{\frac{8k_B T_g}{\pi m_g}}$; at high drift, $\overline{g} \simeq w + \frac{k_B T_g}{m_g w}$. The small-$s$ expansion avoids numerical cancellation.

## S17.3. Poisson event process

Freeze $\nu$ within a step and let $S$ be the probability of no collision. Independent events imply $S(t + dt) = S(t)(1 - \nu dt)$, giving $\frac{dS}{dt} = -\nu S$ in the continuous limit. Hence[16],

$$P(N \geq 1) = 1 - e^{-\nu\Delta t}, \quad P(N = n) = e^{-\lambda}\frac{\lambda^n}{n!}, \quad \lambda = \nu\Delta t. \tag{S43}$$

In explicit mode, each pack undergoes one Bernoulli trial per step and, if selected, one two-body update. The neglected probability of two or more events is $1 - e^{-\lambda}(1 + \lambda) \simeq \frac{\lambda^2}{2}$, approximately $0.468\%$ at $\lambda = 0.1$. The event branch of the joint hybrid model instead samples waiting times $\tau = -\frac{\ln(\xi)}{\nu}$, allowing multiple collisions within one outer step and refreshing the rate after each event. The event rate uses the thermally averaged relative speed. Neutral thermal velocities in an individual IICT update are drawn from an unweighted Maxwell distribution rather than a relative-speed-weighted event distribution.

# S18. Single-collision IICT and energy exchange

## S18.1. Effective pseudo-atom

IICT represents the local ion motion coupled to one collision by an effective pseudo-atom of mass $m_a$[15]. At the specified internal temperature, the pseudo-atom thermal velocity $\boldsymbol{a}$ and neutral thermal velocity $\boldsymbol{g}$ are sampled from isotropic Gaussian distributions:

$$\boldsymbol{a} \sim \mathcal{N}(\boldsymbol{0}, A\boldsymbol{I}), \quad A = \frac{k_B T_i}{m_a}; \quad \boldsymbol{g} \sim \mathcal{N}(\boldsymbol{0}, G\boldsymbol{I}), \quad G = \frac{k_B T_g}{m_g}. \tag{S44}$$

## S18.2. Elastic two-body update

The precollision laboratory-frame velocities are $\boldsymbol{V}_a = \boldsymbol{v} + \boldsymbol{a}$ and $\boldsymbol{V}_g = \boldsymbol{u}_g + \boldsymbol{g}$. The effective head-on collision rule combines momentum conservation with reversal of the relative velocity:

$$m_a \boldsymbol{V}_a + m_g \boldsymbol{V}_g = m_a \boldsymbol{V}_a' + m_g \boldsymbol{V}_g', \quad \boldsymbol{V}_a' - \boldsymbol{V}_g' = -(\boldsymbol{V}_a - \boldsymbol{V}_g). \tag{S45}$$

Solving these equations gives

$$\boldsymbol{V}_a' = \frac{m_a - m_g}{m_a + m_g} \boldsymbol{V}_a + \frac{2m_g}{m_a + m_g} \boldsymbol{V}_g. \tag{S46}$$

Define $b = \frac{2m_g}{m_a + m_g}$, $f = \frac{m_a}{M}$, and $\boldsymbol{\delta} = \boldsymbol{a} + \boldsymbol{w} - \boldsymbol{g}$. Recombining the postcollision pseudo-atom with the rest of the ion by mass weighting gives

$$\boldsymbol{v}' = \boldsymbol{v} + f(\boldsymbol{a} - b\boldsymbol{\delta}). \tag{S47}$$

This follows from $\boldsymbol{v}' = \frac{(M - m_a)\boldsymbol{v} + m_a \boldsymbol{V}_a'}{M}$. The complete-ion velocity increment is thus scaled by the pseudo-atom-to-ion mass ratio $\frac{m_a}{M}$.

## S18.3. Internal-energy update

In IICT, the change in pseudo-atom thermal kinetic energy is evaluated relative to the precollision ion drift[15]:

$$\Delta U = \frac{m_a}{2}\left(|\boldsymbol{a} - b\boldsymbol{\delta}|^2 - |\boldsymbol{a}|^2\right) = \frac{m_a}{2}\left(-2b\boldsymbol{a} \cdot \boldsymbol{\delta} + b^2\ |\boldsymbol{\delta}|^2\right). \tag{S48}$$

The ion internal energy becomes $U' = U + \Delta U$, and the inverse heat-capacity relation gives $T_i'$. IICT-lite requires positive postcollision internal energy. Positive and negative $\Delta U$ correspond to heating and cooling, respectively. Without a collision, the electric field changes center-of-mass translation while internal energy remains fixed; dissociation still accumulates at the current internal temperature.

The pseudo-atom-neutral pair update conserves its momentum and kinetic energy. Molecular internal-energy accounting uses the small-mass-ratio IICT closure, $f = \frac{m_a}{M} \ll 1$, neglecting the recoil correction associated with redefining the molecular center-of-mass frame[15].

### S18.4. Mean heating and cooling

Define $\chi = \frac{4m_a m_g}{(m_a+m_g)^2} = b(2-b)$. The second moments of the independent zero-mean Gaussian velocities give

$$\mathbb{E}[\Delta U] = \frac{3}{2}\chi k_B (T_g - T_i) + \frac{1}{2}\chi m_g \, |\boldsymbol{w}|^2. \tag{S49}$$

The first term is negative when $T_i > T_g$ and describes mean cooling. The second term converts relative drift energy into mean heating. Setting the mean increment to zero gives the local balance temperature at fixed drift,

$$T_{i,*} = T_g + \frac{m_g \, |\boldsymbol{w}|^2}{3k_B}. \tag{S50}$$

The mean velocity increment is $\mathbb{E}[\Delta \boldsymbol{v}] = -c\boldsymbol{w}$, where $c = fb = \frac{2m_a m_g}{M(m_a+m_g)}$. The momentum-relaxation rate is therefore $\gamma_c = \nu c$.

## S19. Conversion between internal energy and temperature in IICT-lite

### S19.1. Heat capacity and effective internal temperature

IICT-lite converts collision energy transfer into effective internal temperature through an invertible relation $U(T_i)$. Internal energy and heat capacity are expressed per ion, in $\frac{\mathrm{J}}{\mathrm{ion}}$ and $\frac{\mathrm{J}}{\mathrm{K \cdot ion}}$, respectively. By definition[6],

$$C_{V(T)} = \frac{dU}{dT}, \quad U(T) = \int_0^T C_{V(\theta)} \, \mathrm{d}\theta, \quad U(0) = 0. \tag{S51}$$

For an independent quantum harmonic mode of frequency $\nu_l$, the mean thermal energy above the zero-point level and its heat capacity are

$$U_{l(T)} = \frac{h\nu_l}{e^{x_l} - 1}, \quad C_{V,l}(T) = k_B \frac{x_l^2 e^{x_l}}{(e^{x_l} - 1)^2}, \quad x_l = \frac{h\nu_l}{k_B T}. \tag{S52}$$

When $x_l \ll 1$, each mode contributes $k_B$ to the heat capacity; at lower temperatures, high-frequency modes contribute less. IICT-lite provides classical vibrational, constant-heat-capacity, and tabulated models.

## S19.2. Classical vibrational and constant-heat-capacity models

A nonlinear molecule with $N$ atoms has $d_v = 3N - 6$ vibrational degrees of freedom. In the classical model, each mode contributes $k_B$:

$$C_V = d_v k_B, \quad U(T_i) = d_v k_B T_i, \quad T_{i(U)} = \frac{U}{d_v k_B}. \tag{S53}$$

The atom count sets $d_v$, corresponding to the classical limit of fully excited vibrational modes. The constant-heat-capacity model instead uses a user-specified value $C_0 > 0$ per ion:

$$C_V = C_0, \quad U(T_i) = C_0 T_i, \quad T_{i(U)} = \frac{U}{C_0}. \tag{S54}$$

Both Eq. 53 and Eq. 54 use $U(0) = 0$. A larger heat capacity produces a smaller temperature change for the same energy increment, with the local approximation $\Delta T_i \simeq \Delta \frac{U}{C_{V(T_i)}}$.

## S19.3. Tabulated model and inversion

The tabulated model accepts CSV data as either $(T_j, U_j)$ or $(T_j, C_{V,j})$. Temperature nodes must increase strictly. Energy nodes must also increase strictly, or heat-capacity values must be positive, ensuring an invertible mapping. Input units are K, J/ion, and J/(K ion), respectively.

For a heat-capacity table, energy nodes are constructed from zero temperature using the trapezoidal rule:

$$U_0 = 0, \quad U_{j+1} = U_j + \frac{C_{V,j} + C_{V,j+1}}{2}(T_{j+1} - T_j). \tag{S55}$$

If the first input temperature is positive, a $T_0 = 0$ node is prepended with the first supplied heat capacity. A directly supplied energy table is similarly extended by $(0, 0)$ when needed; an existing zero-temperature node must have zero energy.

Runtime conversion uses piecewise-linear interpolation of the energy nodes. Exchanging the interpolation axes gives the inverse:

$$\begin{aligned} U(T) &= U_j + \frac{T - T_j}{T_{j+1} - T_j}(U_{j+1} - U_j), \\ T(U) &= T_j + \frac{U - U_j}{U_{j+1} - U_j}(T_{j+1} - T_j). \end{aligned} \tag{S56}$$

Outside the table, Eq. 56 is extended linearly using the nearest endpoint slope, with finite, nonnegative results required. Each collision updates $U' = U + \Delta U > 0$, followed by inversion to $T_i'$. Heat capacity determines the temperature increase associated with energy transfer, whereas the activation enthalpy and entropy in Section 15 determine the dissociation rate at that temperature.

## S20. Pseudo-atom mass parameterization in IICT-lite

### S20.1. Constant-mass model

The pseudo-atom mass $m_a$ is the effective mass coupling a collision to internal ion motion. IICT-lite can use a user-specified constant $m_0$:

$$m_{a(T_i,w)} = m_0, \quad w = |\boldsymbol{v} - \boldsymbol{u}_g|, \quad 0 < m_0 \leq M. \tag{S57}$$

The input mass is in Da and is multiplied by the atomic mass constant $m_u$ to obtain kg. This model assigns one coupling mass to the ion-gas combination under study.

### S20.2. Temperature-speed table

A temperature- and drift-dependent coupling mass can be supplied as a complete rectangular grid of $T_i$, $w$, and $m_a$. For neighboring nodes $T_j, T_{j+1}$ and $w_k, w_{k+1}$, define $\alpha = \frac{T_i - T_j}{T_{j+1} - T_j}$ and $\beta = \frac{w - w_k}{w_{k+1} - w_k}$. Bilinear interpolation gives

$$\begin{aligned} m_{a(T_i,w)} &= (1-\alpha)(1-\beta)m_{j,k} + \alpha(1-\beta)m_{j+1,k} \\ &\quad + (1-\alpha)\beta m_{j,k+1} + \alpha\beta m_{j+1,k+1}. \end{aligned} \tag{S58}$$

Temperature, speed, and tabulated mass are supplied in K, m/s, and Da, respectively. Queries outside the table use the nearest boundary coordinate. Constant and tabulated models enforce the configured mass bounds; the collision core also requires $0 < m_a \leq M$. A parameter set may be associated with a gas name to distinguish collision partners.

### S20.3. Effect on momentum and energy exchange

The mean momentum and internal-energy updates in Section 11 contain the factors

$$c = \frac{2m_a m_g}{M(m_a + m_g)}, \quad \chi = \frac{4m_a m_g}{(m_a + m_g)^2}. \tag{S59}$$

At fixed $M$ and $m_g$, Eq. 59 shows that $c$ increases with $m_a$, whereas $\chi$ reaches its maximum of 1 at $m_a = m_g$. Pseudo-atom mass therefore affects both momentum relaxation and collisional heating or cooling. The effective cross section controls event frequency through $\nu$ in Section 10.

IICT-lite resolves heat capacity, pseudo-atom mass, and dissociation kinetics as independent parameter modules. Parameters can be supplied through JSON files, CSV tables, and command-line overrides. Reports retain the effective values, their sources, and the parameter-configuration SHA-256 for reproducibility.

## S21. Hybrid Langevin model

### S21.1. Moments of a single collision jump

Let $Z = \left(v_x, v_y, v_z, U\right)$. Using the IICT definitions of $b$, $f$, and $c$, write

$$\boldsymbol{h} = (1-b)\boldsymbol{a} + b\boldsymbol{g}, \quad S = (1-b)^2 A + b^2 G. \tag{S60}$$

Then $\Delta \boldsymbol{v} = f\boldsymbol{h} - c\boldsymbol{w}$ and $\Delta U = \frac{1}{2} m_{a(|\boldsymbol{h}-b\boldsymbol{w}|^2 - |\boldsymbol{a}|^2)}$. With $\mathbb{E}[\boldsymbol{h}] = 0$, $\mathbb{E}\left[h_i h_j\right] = S\delta_{ij}$, and $\mathbb{E}\left[h_i a_j\right] = (1-b)A\delta_{ij}$,

$$\boldsymbol{\mu}_v = -c\boldsymbol{w}, \quad \mu_U = \frac{m_a}{2}\left[3(S-A) + b^2\ |\boldsymbol{w}|^2\right], \tag{S61}$$

$$\mathbb{E}[\Delta \boldsymbol{v} \Delta \boldsymbol{v}^\top] = f^2 S \boldsymbol{I} + c^2 \boldsymbol{w}\boldsymbol{w}^\top, \tag{S62}$$

$$\mathbb{E}[\Delta \boldsymbol{v} \Delta U] = (-c\boldsymbol{w})\mu_U - f m_a b S \boldsymbol{w}. \tag{S63}$$

For Gaussian variables, $\mathbb{E}[X^4] = 3\mathbb{E}[X^2]^2$. Thus $\mathrm{Var}(|\boldsymbol{h}|^2) = 6S^2$ and $\mathrm{Cov}(|\boldsymbol{h}|^2, |\boldsymbol{a}|^2) = 6(1-b)^2 A^2$. Third-order cross moments between the linear and centered quadratic terms vanish, giving

$$\mathrm{Var}(\Delta U) = \frac{3}{2} m_a^2 \left[(S-A)^2 + 2Ab^2 G\right] + m_a^2 b^2 S\ |\boldsymbol{w}|^2, \tag{S64}$$

$$\mathbb{E}[(\Delta U)^2] = \mathrm{Var}(\Delta U) + \mu_U^2. \tag{S65}$$

### S21.2. Diffusion approximation of the jump process

During a short interval $dt$, the number of jumps is approximately Poisson with mean $\nu dt$. If $\Delta Z$ is the increment conditional on one jump, the compound-Poisson moments are[16]

$$\mathbb{E}[dZ] = \nu \mathbb{E}[\Delta Z] dt, \quad \mathrm{Cov}(dZ) = \nu \mathbb{E}[\Delta Z \Delta Z^\top] dt + O(dt^2). \tag{S66}$$

The law of total covariance includes the outer product of the mean jump because the event count fluctuates. The diffusion coefficient therefore uses the raw second moment of the single-jump increment. Truncating the Kramers-Moyal expansion after the second term gives the Ito diffusion approximation,

$$dZ = \nu \mathbb{E}[\Delta Z] dt + B_Z d\boldsymbol{W}_t, \quad B_Z B_Z^\top = \nu \mathbb{E}[\Delta Z \Delta Z^\top]. \tag{S67}$$

This approximation requires sufficiently many small jumps within the aggregation interval.

### S21.3. Exponential integration with frozen coefficients

Within an internal substep, $\nu$, the local moments, and the gas state are frozen. The velocity drift remains linear, $-\gamma_c \boldsymbol{w}$, with $\gamma_c = \nu c$. Integrating the resulting linear stochastic equation gives[16]

$$\boldsymbol{w}' = e^{-\gamma_c \tau} \boldsymbol{w} + \boldsymbol{\eta}_v, \quad U' = U + \nu \mu_U \tau + \eta_U. \tag{S68}$$

Let $Q_{vv}$, $Q_{vU}$, and $Q_{UU}$ denote the single-jump raw second moments above. Ito isometry gives the integrated noise covariances:

$$\Sigma_{vv} = \frac{\nu(1 - e^{-2\gamma_c \tau})}{2\gamma_c} Q_{vv}, \quad \Sigma_{vU} = \frac{\nu(1 - e^{-\gamma_c \tau})}{\gamma_c} Q_{vU}, \quad \Sigma_{UU} = \nu\tau Q_{UU}. \tag{S69}$$

As $\gamma_c \to 0$, the exponential ratios tend continuously to $\tau$. Velocity noise is generated from its isotropic and drift-aligned rank-one components. A conditional Gaussian construction then supplies correlated internal-energy noise:

$$\eta_U = \Sigma_{Uv}\Sigma_{vv}^{-1}\boldsymbol{\eta}_v + \sqrt{\Sigma_{UU} - \Sigma_{Uv}\Sigma_{vv}^{-1}\Sigma_{vU}}\,\xi_U. \tag{S70}$$

The update preserves velocity-energy noise correlations and integrates the frozen-coefficient linear stochastic system exactly. Constraints such as $\lambda c \le c_{\mathrm{relax}}$ limit the change in local state within a substep.

### S21.4. Aggregation criteria and event fallback

Let $\lambda = \nu\tau$ and let $dU$ be the distance to the nearest allowed internal-energy boundary. The algorithm limits mean energy drift, its standard deviation, and total momentum relaxation:

$$\lambda c \le c_{\mathrm{relax}}, \quad \lambda\,|\mu_U| \le \min\left(c_U U, \frac{dU}{2}\right), \quad \sqrt{\lambda Q_{UU}} \le \min\left(c_U U, \frac{dU}{16}\right). \tag{S71}$$

Aggregation also requires $\lambda \ge N_{\mathrm{min}}$. The defaults are $N_{\mathrm{min}} = 8$, $c_{\mathrm{relax}} = 0.1$, and $c_U = 0.05$. The energy constraints reserve boundary margins for the mean drift and eight standard deviations. Eligible substeps use the joint diffusion update; otherwise, two-body events are processed individually with exponentially distributed waiting times.

## S22. Dissociation kinetics and precursor survival

### S22.1. Transition-state rate

Transition-state theory treats reactants and activated complexes as being in approximate quasi-equilibrium, with crossing frequency $(k_B T)/h$[17]. Taking the transmission coefficient $\kappa = 1$ and $\Delta G^\ddagger = \Delta H^\ddagger - T\Delta S^\ddagger$ gives

$$k(T_i) = \frac{k_B T_i}{h}\exp\left(-\frac{\Delta G^\ddagger}{RT_i}\right) = \frac{k_B T_i}{h}\exp\left(\frac{\Delta S^\ddagger}{R} - \frac{\Delta H^\ddagger}{RT_i}\right). \tag{S72}$$

This canonical parameterization assumes rapid redistribution of internal energy and uses effective internal temperature with specified activation parameters to describe precursor loss. Published thermochemical analyses of protonated leucine enkephalin provide a source of activation parameters for this rate expression[4].

For constant activation enthalpy and entropy, the rate varies as $T\exp\left[-\Delta\frac{H^\ddagger}{RT}\right]$. Its logarithmic temperature sensitivity is

$$\frac{d\ln k}{dT} = \frac{1}{T} + \frac{\Delta H^\ddagger}{RT^2}. \tag{S73}$$

The corresponding local Arrhenius activation energy is $E_a = \Delta H^\ddagger + RT$. Increasing activation enthalpy lowers the rate, while increasing activation entropy raises it.

## S22.2. From first-order loss to cumulative hazard

Along a trajectory, the represented precursor weight satisfies $\frac{dW_p}{dt} = -k\left(T_{i(t)}\right)W_p$. Separation and integration give

$$\mathcal{H}(t) = \int_{t_0}^{t} k(T_i(\tau))\,\mathrm{d}\tau, \quad S(t) = \frac{W_p(t)}{W_p(t_0)} = e^{-\mathcal{H}(t)}, \quad P_{\text{diss}} = 1 - S. \tag{S74}$$

Dissociation depends on the complete temperature-time history. A short high-temperature interval can contribute most of the final loss because $k(T)$ is large. At constant temperature, the half-life is $t_{\frac{1}{2}} = \frac{\ln(2)}{k(T)}$.

Using the postcollision state for the micro-step rate integral gives

$$\Delta\mathcal{H} = k\left(T_{i,\ \text{post}}\right)\Delta t_{\text{alive}}, \quad W_p^{n+1} = W_p^n e^{-\Delta\mathcal{H}}, \quad \Delta W_d = W_p^n\left(1 - e^{-\Delta\mathcal{H}}\right). \tag{S75}$$

The update first evaluates the hazard increment, $k\Delta t_{\text{alive}}$, and then converts it to precursor loss through the exponential survival relation. Small increments are evaluated with `expm1` to reduce cancellation.

The actual exposure time is

$$\Delta t_{\text{alive}} = \max\{0, \min(t_n + \Delta t, t_{\text{terminal}}) - \max(t_n, t_{\text{birth}})\}. \tag{S76}$$

Hazard accumulates over each active ion's physical lifetime, including micro steps without explicit collisions. An unchanged temperature allows reuse of its rate constant, while hazard continues to accumulate at every step.

## S22.3. Transport and loss modes

In `transport` mode, transport weight $W$ and PIC charge $qW$ remain fixed. Precursor weight $W_p$ and dissociation-tagged weight $W_d = W - W_p$ are recorded separately, using the original precursor transport parameters and trajectory. The outputs are predicted dissociated precursor fraction and survival yield.

In `loss` mode, transport weight and its associated charge decrease with precursor loss. The slot is released when the precursor is exhausted, and dissociation loss is accounted for separately. The `off` mode disables dissociation.

Effective internal temperature remains defined by the selected precursor energy model, and dissociation is accumulated as a kinetic observable along the transport trajectory.

# S23. Electrode and terminal-event localization

## S23.1. Segment approximation

Within each micro step, the path between its endpoints is approximated by a line segment:

$$\boldsymbol{x}(\alpha) = \boldsymbol{x}_0 + \alpha(\boldsymbol{x}_1 - \boldsymbol{x}_0), \quad 0 \le \alpha \le 1. \tag{S77}$$

This approximation and the geometric CFL constraint are used to locate boundary crossings.
For a detector plane $z = z_d$, the intersection parameter follows directly:

$$\alpha_z = \frac{z_d - z_0}{z_1 - z_0}, \quad t_{\text{hit}} = t_0 + \alpha_z \Delta t. \tag{S78}$$

Only crossings in the specified direction with a parameter in $[0, 1]$ are accepted. Events within the detector radius are classified as `z_exit` those outside it are `radial_out`.

## S23.2. Radial and domain boundaries

Let $\boldsymbol{d} = \boldsymbol{x}_1 - \boldsymbol{x}_0$. Expanding $x(\alpha)^2 + y(\alpha)^2 = R^2$ gives

$$\left(d_x^2 + d_y^2\right)\alpha^2 + 2\left(x_0 d_x + y_0 d_y\right)\alpha + \left(x_0^2 + y_0^2 - R^2\right) = 0. \tag{S79}$$

The first valid outward crossing is selected. Upstream and downstream domain-plane crossings are treated similarly and classified as `domain_out`. Electrode detection uses the metal mask and a surface-distance threshold. Sampling along the segment brackets the first triggering interval, which is refined by bisection. After $m$ bisections, the parameter interval is reduced by $2^{-m}$; the mask resolution sets the underlying geometric accuracy.
The earliest candidate event is selected. Ties follow an explicit priority order: electrode, radial boundary, exit plane, then domain boundary. Position, velocity, and time are interpolated within the segment. The event time also truncates dissociation exposure at the terminal crossing.
Electrodes are absorbing boundaries. The surface-distance threshold and mask resolution determine the precision of impact localization.

# S24. Exit current and statistical observables

## S24.1. Current and transmission

Within a time window $[t_a, t_b)$, the exit current is the charge carried through the detector plane divided by the window duration:

$$I_{\text{exit}} = \frac{1}{t_b - t_a} \sum_{j \in \mathcal{E}} W_j z_{q,j} e. \tag{S80}$$

This is positive for a single positive charge state. Current magnitudes use $|q|$. The steady flux transmission is $\eta_I = \frac{I_{\text{exit}}}{I_{\text{in}}}$, whereas the cumulative transmitted fraction is $\eta_N = \frac{\sum W_{\text{exit}}}{\sum W_{\text{injected}}}$. They differ during startup because part of the injected population remains inside the domain.

For each loss category, $I_s = \frac{|q| \sum_s W}{\Delta t}$ and $f_s = \frac{I_s}{I_{\text{in}}}$ give the input-flux fractions lost to electrodes, domain boundaries, or the radial limit. Terminal flux fractions in a finite window can differ from a total of 100% because the domain inventory changes. Conservation requires

$$N_{\text{active}}(t_b) - N_{\text{active}}(t_a) = N_{\text{injected,window}} - \sum_s N_{s,\text{ window}}. \tag{S81}$$

The inventory balance accounts consistently for buffered ions, admitted source injection, and dissociation loss in `loss` mode. In `transport` mode, dissociation tagging is a trajectory attribute that may coexist with an exit or electrode terminal category.

## S24.2. Precursor survival and experimental fragment fraction

Predicted exit survival and dissociation percentage are normalized by transport weight:

$$\text{SY}_{\text{exit}} = \frac{\sum_{j \in \mathcal{E}} W_{p,j}}{\sum_{j \in \mathcal{E}} W_j}, \quad F_{\text{diss,exit}} = 100(1 - \text{SY}_{\text{exit}})\%. \tag{S82}$$

These observables describe the transmitted population. The experimental ratio $\frac{\sum I_{\text{fragment}}}{I_{\text{precursor}} + \sum I_{\text{fragment}}}$ is a signal fraction. Comparison with simulation also depends on fragment selection and species-dependent transmission and detector response.

## S24.3. Kinetic energy, temperature, and weighted quantiles

The laboratory-frame translational kinetic energy of an exit event is $K = \frac{M|\boldsymbol{v}|^2}{2}$, converted to eV by division by the elementary charge. It includes directed drift. Effective internal temperature is calculated separately from molecular internal energy.

The weighted mean and empirical cumulative distribution are

$$\langle X \rangle_W = \frac{\sum_j W_j X_j}{\sum_j W_j}, \quad F_W(x) = \frac{\sum_j W_j \mathbf{1}(X_j \le x)}{\sum_j W_j}. \tag{S83}$$

The quantile $Q_p = \inf\left\{x : F_{W(x)} \ge p\right\}$ is obtained by sorting the samples and selecting the first cumulative weight reaching the threshold. The median is $Q_{0.5}$, and the central 90% interval is $[Q_{0.05}, Q_{0.95}]$, describing the population spread.

### S24.4. Beam radius and residence time

For exit radii $r_j = \sqrt{x_j^2 + y_j^2}$, the weighted 95% quantile $r_{95}$ is the radius of an axis-centered circle containing 95% of the represented exit population. The root-mean-square radius is $r_{\mathrm{rms}} = \sqrt{\langle r^2 \rangle_W}$.

Time of flight is the terminal time minus the birth time at activation into the transport region. Each pack has a unique track ID for following its position, internal temperature, and precursor survival across snapshots.

### S24.5. Steady-state selection and uncertainty

Steady-state windows are selected from consecutive-window behavior of exit current, survival yield, kinetic energy, internal temperature, and domain inventory. The binned-current coefficient of variation, $\mathrm{CV} = \frac{\mathrm{std}(I_{\mathrm{bin}})}{\mathrm{mean}(I_{\mathrm{bin}})}$, is reported together with changes between earlier and later window means.

For independent weighted samples, the effective sample size is $N_{\mathrm{eff}} = \frac{(\sum W)^2}{\sum W^2}$[18]. This follows by comparing the weighted-mean variance, $\mathrm{Var}\left(\overline{X}\right) = \frac{\sigma_X^2 \sum W^2}{(\sum W)^2}$, with its equal-weight counterpart. Independent random seeds or time-block statistics can additionally assess uncertainty in transport data with shared-field and temporal correlations.

## Supplementary references

(1) Paris, L. R.; Green, A. W.; Prell, J. S. Computed Vibrational Heat Capacities for Gas-Phase Biomolecular Ions. *Journal of the American Society for Mass Spectrometry* **2025**, *36* (4), 862.

(2) Zhou, Z.; Luo, M.; Chen, X.; Yin, Y.; Xiong, X.; Wang, R.; Zhu, Z.-J. Ion Mobility Collision Cross-Section Atlas for Known and Unknown Metabolite Annotation in Untargeted Metabolomics. *Nature communications* **2020**, *11* (1), 4334.

(3) Nichols, C. M.; May, J. C.; Sherrod, S. D.; McLean, J. A. Automated Flow Injection Method for the High Precision Determination of Drift Tube Ion Mobility Collision Cross Sections. *Analyst* **2018**, *143* (7), 1556–1559.

(4) Sztáray, J.; Memboeuf, A.; Drahos, L.; Vékey, K. Leucine Enkephalin—a Mass Spectrometry Standard. *Mass spectrometry reviews* **2011**, *30* (2), 298–320.

(5) Hockney, R. W.; Eastwood, J. W. *Computer Simulation Using Particles*; CRC Press, 1988.

(6) McQuarrie, D. A. *Statistical Mechanics*; University Science Books, 2000.

(7) Eymard, R.; Gallouët, T.; Herbin, R. Finite Volume Approximation of Elliptic Problems and Convergence of an Approximate Gradient. *Applied Numerical Mathematics* **2001**, *37* (1–2), 31–53.

(8) Hestenes, M. R.; Stiefel, E. Methods of Conjugate Gradients for Solving Linear Systems. *Journal of Research of the National Bureau of Standards* **1952**, *49* (6), 409–436.

(9) Ruge, J. W.; Stüben, K. Algebraic Multigrid. In *Multigrid Methods*; McCormick, S. F., Ed.; Frontiers in Applied Mathematics; Society for Industrial, Applied Mathematics, 1987; Vol. 3, pp 73–130.

(10) Bell, N.; Olson, L. N.; Schroder, J.; Southworth, B. PyAMG: Algebraic Multigrid Solvers in Python. *Journal of Open Source Software* **2023**, *8* (87), 5495.

(11) Young, D. Iterative Methods for Solving Partial Difference Equations of Elliptic Type. *Transactions of the American Mathematical Society* **1954**, *76* (1), 92–111.

(12) Gerlich, D. Inhomogeneous RF Fields: A Versatile Tool for the Study of Processes with Slow Ions. In *Advances in Chemical Physics*; John Wiley & Sons, 1992; Vol. 82, pp 1–176.

(13) Butcher, J. C. *Numerical Methods for Ordinary Differential Equations*, 3rd ed.; John Wiley & Sons, 2016.

(14) Strang, G. On the Construction and Comparison of Difference Schemes. *SIAM Journal on Numerical Analysis* **1968**, *5* (3), 506–517.

(15) Prell, J. S. Modeling Collisional Kinetic Energy Damping, Heating, And Cooling of Ions in Mass Spectrometers: A Tutorial Perspective. *International Journal of Mass Spectrometry* **2024**, *504*, 117290.

(16) Gardiner, C. *Stochastic Methods: A Handbook for the Natural and Social Sciences*, 4th ed.; Springer, 2009.

(17) Eyring, H. The Activated Complex in Chemical Reactions. *The Journal of Chemical Physics* **1935**, *3* (2), 107–115.

(18) Kish, L. *Survey Sampling*; Wiley Classics Library; John Wiley & Sons, 1995.